\documentclass[nonacm,sigplan,10pt]{acmart}

\usepackage[utf8]{inputenc}
\usepackage{epsfig,xspace,xspace,multirow}
\usepackage[labelfont=bf,textfont={small, it}]{caption}
\usepackage{parskip}
\usepackage{tabu}
\usepackage[normalem]{ulem}
\usepackage{soul}
\usepackage{hyperref}
\usepackage{array}
\usepackage{esvect}
\usepackage{textcomp}
\usepackage{amsmath}
\usepackage{booktabs}
\usepackage{siunitx}
\usepackage[T1]{fontenc}
\usepackage{makecell}
\usepackage{setspace}
\usepackage[hang,flushmargin]{footmisc} 
\usepackage{enumitem}

\usepackage{tikz}
\usepackage{diagbox}
\usepackage{tcolorbox}
\usepackage{color}
\usepackage{flushend}
\usepackage{graphicx}
\usepackage{listings}
\usepackage[linesnumbered, ruled, lined, noend]{algorithm2e}
\usepackage{xurl}
\usepackage{subcaption}
\usepackage{caption}

\definecolor{cmured}{RGB}{176, 28, 51}

\newif\ifsubmitvar
\submitvartrue

\ifsubmitvar
    \newcommand{\brandon}[1]{}
    \newcommand{\zhuo}[1]{}
    \newcommand{\change}[1]{#1}
\else
    \newcommand{\brandon}[1]{{\color{cmured} {(brandon: #1)}}}
    \newcommand{\zhuo}[1]{{\color{blue} {(zhuo: #1)}}}
    \newcommand{\change}[1]{{\color{orange} {(change: #1)}}}
\fi

\newcommand{\maybedelete}[1]{}
\newcommand{\mydelete}[1]{}

\newcounter{packednmbr}

\newcommand{\mybox}[1]{\fbox{\begin{minipage}{0.97\linewidth}{#1}\end{minipage}}}

\newcommand{\mypara}[1]{\myparatight{#1}~}

\newcommand{\myparatight}[1]{\vspace{-6pt}\smallskip\noindent{\bf {#1}}~}

\newenvironment{packeditemize}{\begin{list}{$\bullet$}{\setlength{\itemsep}{0.5pt}\addtolength{\labelwidth}{-4pt}\setlength{\leftmargin}{\labelwidth}\setlength{\listparindent}{\parindent}\setlength{\parsep}{1pt}\setlength{\topsep}{0pt}}}{\end{list}}

\def\Snospace~{\S{}}

\newcommand{\ie}{\emph{i.e.,}\xspace}
\newcommand{\eg}{\emph{e.g.,}\xspace}

\newcommand{\mytextapprox}{\raisebox{0.5ex}{\texttildelow}}

\newcommand{\SYS}{\textsc{Gyrfalcon}\xspace}
\newcommand{\sys}{\textsc{Gyrfalcon}\xspace}

\newcommand{\remove}[1]{}

\definecolor{darkgreen}{RGB}{73, 175, 73}

\newcommand{\mydeg}{$^{\circ}$}

\newcommand{\SampledLLDD}{\textit{Sampled L2D2}}

\newcommand{\naiveCaptureRatio}{more than $90\%$}

\newcommand{\evalEtwoESYSoverLLDD}{$5.6\text{--}8.2\times$}

\newcommand{\evalIncCapture}{$1.6\text{--}2.2\times$}
\newcommand{\evalIncTX}{$1.5\times$}

\newcommand{\evalMorePlaneImprove}{$7.9\text{--}10.5\times$}

 \usepackage[compact]{titlesec}

\hypersetup{pdfstartview=FitH,pdfpagelayout=SinglePage}
\begin{document}

\title{
Nanosatellite Constellation and Ground Station Co-design for Low-Latency Critical Event Detection
}

\author{Zhuo Cheng \hspace{1.5cm} Brandon Lucia}
\affiliation{Carnegie Mellon University \country{}}

\date{}

\begin{abstract}

Advancements in nanosatellite technology lead to more Earth-observation satellites in low-Earth orbit.
We explore using nanosatellite constellations to achieve low-latency detection for time-critical events, such as forest fires, oil spills, and floods.
The detection latency comprises three parts: capture, compute and
transmission.
Previous solutions reduce transmission latency, but we find that the bottleneck
is capture latency, accounting for $>$90\% of end-to-end latency.
We present a measurement study on how various satellite and ground station design factors affect latency. We offer design guidance to operators on how to choose satellite orbital configurations and design an algorithm to choose ground station locations. For six use cases, our design guidance reduces end-to-end latency by \evalEtwoESYSoverLLDD\ compared to the existing system.

\end{abstract}

\setcopyright{none}
\settopmatter{printacmref=false} 
\renewcommand\footnotetextcopyrightpermission[1]{} 
\pagestyle{plain}

\settopmatter{printfolios=true}

\maketitle

\section{Introduction}

Recent advances in machine learning have enabled the analysis of Earth
observation data for critical applications, including forest fire
detection~\cite{fire-detect-multiple-kernel, fire-sat-dataset}, oil spill
monitoring~\cite{nasa-satellite-image-oil-spill, oil-spill-mdpi2019}, and flood
detection~\cite{sat-flood-detect-22-mdpi, sat-flood-detect-20-china-mdpi}.
However, previous approaches downlink and process all data on Earth, incurring
long, unpredictable {\em latency}, which can be tens of hours or days.
%
Moreover, legacy missions use few satellites because each is extremely costly,
resulting in limited coverage and potential event misses; concretely, NASA
Landsat 9 costs around \$750 million~\cite{landsat9-cost} and takes 16 days to
fully image Earth.

%



\begin{figure}[t]
    \centering
    \begin{subfigure}[t]{0.32\linewidth}
        \centering
        \includegraphics[width=0.95\linewidth]{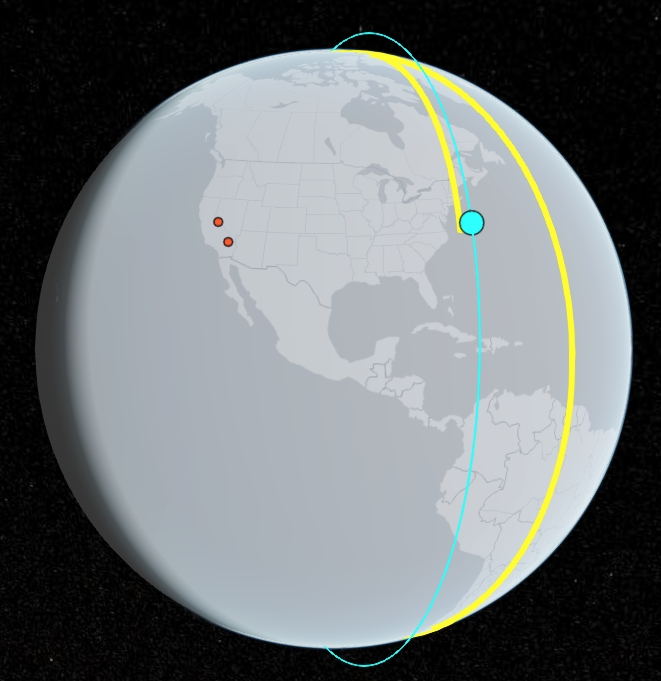}
        \caption{Two fires start in California}
    \end{subfigure}
    \begin{subfigure}[t]{0.32\linewidth}
        \centering
        \includegraphics[width=0.95\linewidth]{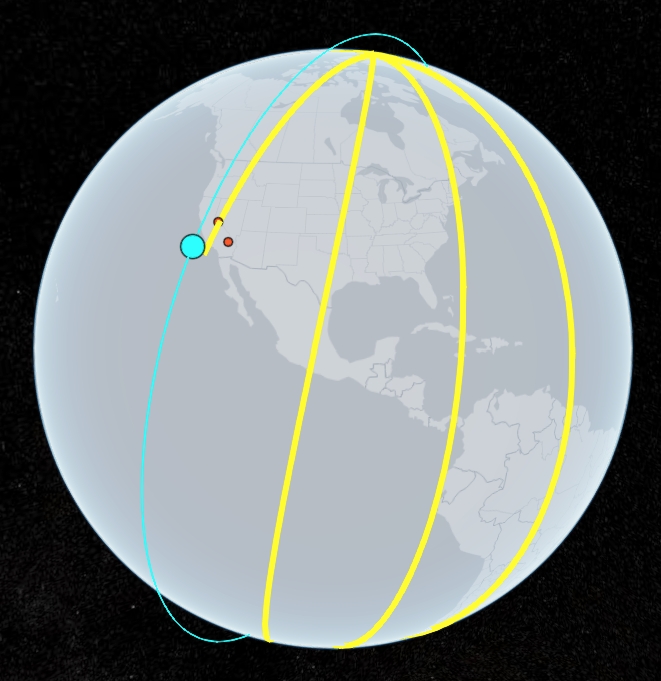}
        \caption{Capture}
    \end{subfigure}
    \begin{subfigure}[t]{0.32\linewidth}
        \centering
        \includegraphics[width=0.95\linewidth]{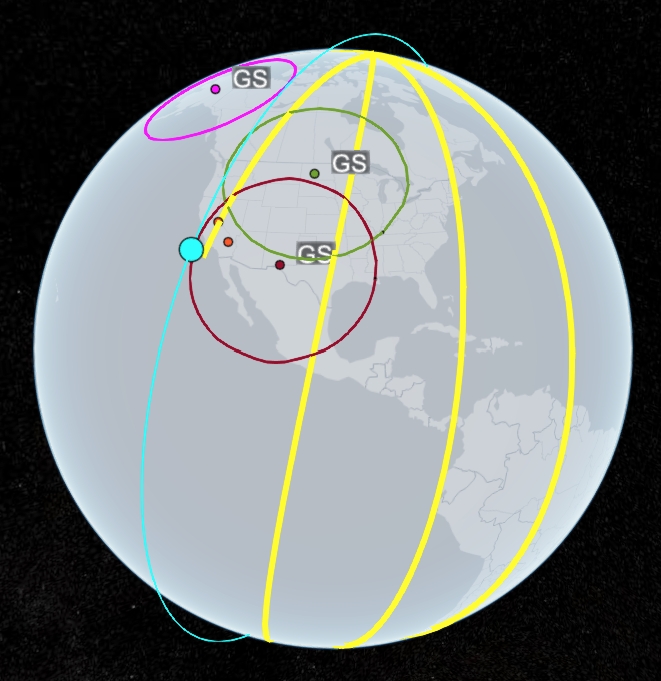}
        \caption{Transmission}
    \end{subfigure}
\caption{
    Event detection involves three latency components: capture, compute (not shown) transmission. 
    Blue dot: satellite, blue line: satellite's orbit, yellow line: satellite's ground track.
\label{fig:design:sys_model}
    } 
\end{figure}

\begin{figure}[t]
    \centering
    \centering
    \includegraphics[width=0.6\linewidth]{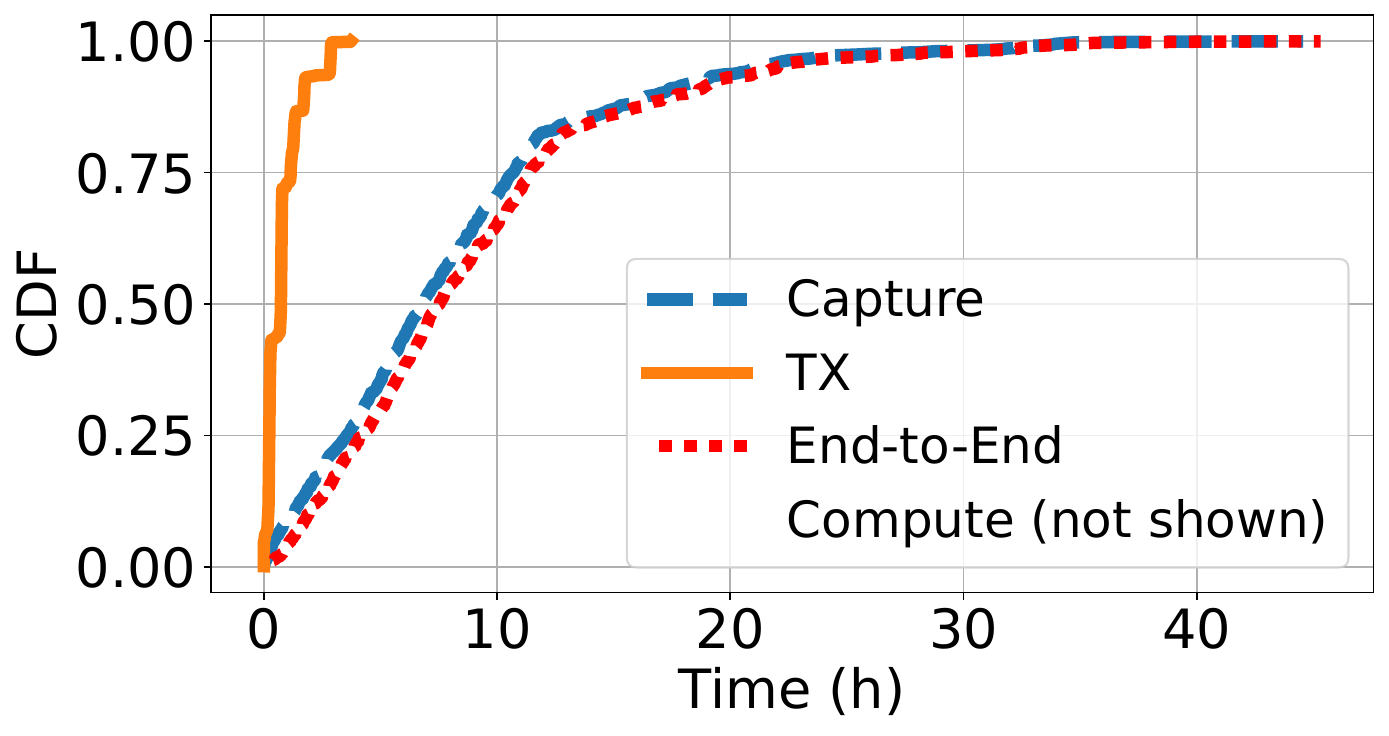}
    \caption{
    In the existing Planet constellation, which comprises 160 satellites and 12 ground stations, the primary latency bottleneck is the capture latency, responsible for \naiveCaptureRatio\ of the total end-to-end latency.
    \label{fig:motivation:planet_baseline}
    } 
\end{figure}

Two recent technological shifts provide new opportunities to improve coverage and reduce latency.
The first shift is the emergence of low-cost nanosatellites.  For example, a
``1U'' CubeSat measures $10cm \times 10cm \times 10cm$ and weighs a few
kilograms. CubeSat manufacturing and launch costs total to a few ten thousand
dollars.
The low cost allows large {\em constellations}, such as Planet's 160 Dove
nanosatellites, expanding coverage.
The second shift is the advent of orbital edge computing~\cite{sat-cal2019, oec-asplos20, kodan-asplos23}, where satellites process images on-board instead of sending raw data to Earth for processing. 
This provides opportunities for real-time decision-making in sophisticated Earth observation applications.

Given these new opportunities, we consider whether we can provide high-coverage, low-latency detection at a low cost. 
%
\autoref{fig:design:sys_model} shows the end-to-end latency for a critical event, which is the timespan between when an event begins (\eg fires start in California) and when a satellite transmits an alert about the event to the ground. End-to-end latency has three components: capture latency (the time for a satellite in the constellation to pass over the event region, typically several hours), computation latency (the time for the satellite to process the image and identify the event, typically a few seconds, not shown), and transmission latency (the time for the satellite to pass over a ground station, typically tens of minutes).
We assume that with orbital edge computing, only a small fraction of images need to be downloaded for critical event detection, as such events are infrequent. Consequently, the transmission latency is primarily determined by the time required for a satellite to establish a connection with the ground station, rather than by the communication bandwidth, which is different the case considered by L2D2~\cite{l2d2-sigcomm}, Umbra~\cite{deepak-umbra-mobicom23}. Additionally, we later demonstrate that the satellite possesses sufficient computational power to handle image processing without causing queuing delays.


Existing constellations fail to achieve high-coverage, low-latency, or both for critical event detection applications.
For example, the Dove~\cite{Planet-dove-constellation} constellation, operated by Planet, is one of the largest cubesat Earth observation constellations, with 160 satellites and 12 ground stations.
We run simulations based on historical fire locations to determine the latency of the Planet deployment in detecting modeled wildfires.
As \autoref{fig:motivation:planet_baseline} shows, 
the median end-to-end latency is 7.5 hours, which is far too long for an operator to be responsive to a critical event.  Moreover, the simulations show that the  
capture latency is the bottleneck, accounting for \naiveCaptureRatio\ of the end-to-end latency.
Existing work~\cite{l2d2-sigcomm, deepak-umbra-mobicom23, deepak-nsdi24-serval,OrbitCast-icpn21-zeqi} on nanosatellites focuses on reducing transmission latency without addressing capture latency.
Also, these solutions rely on large numbers of ground stations or high complexity communication satellites, incurring substantial costs. 
%


In this work, we first identify the design components of satellite and ground stations and then present a measurement study on how each factor affects latency.
The satellite design encompasses decisions regarding hardware components (\eg sensors, computers, solar panels) and orbital parameters (\eg inclination and number of planes). 
The ground station design entails choices related to hardware components (\eg antennas, dishes) and their respective locations.
The selection of satellite and ground station hardware components is primarily based on mission-specific requirements.
On the other hand, orbital parameters decide the coverage area of the constellation, directly impacting capture latency.
Ground station locations determines the timing of the satellite-ground communication sessions, directly impacting transmission latency.
Therefore, we focus on
the design of the constellation's orbital configuration and ground station locations.

\change{
We consider several use cases: wildfire detection, earthquake signature detection, city monitoring, and wildlife \& forest monitoring. Using historical event locations, we conduct a simulation-based measurement study to evaluate latency by simulating satellite trajectories over time.

For constellation design, we provide a detailed measurement study on how different satellite orbital parameters affect capture latency and transmission latency. 
For ground station design,
we provide an algorithm to determine the optimal ground station locations to minimize transmission latency.

Our key findings are as follow:

\begin{packeditemize}
\item 
Deploying satellites in multiple planes, rather than a single plane, significantly reduces capture latency. The existing Planet constellation does not optimize for latency and places all satellites in the same plane. We find that using 10 planes reduces capture latency by \evalMorePlaneImprove.

\item 
Communication constellations like Starlink use a lower 53-degree inclination orbit so satellites spend more time at middle latitudes, providing better services where most customers live. However, we find that lower inclinations minimally reduce latency for Earth observation tasks.

\item
Prior work~\cite{l2d2-sigcomm} proposes using geo-distributed ground stations to reduce transmission latency. We find that naively selecting geo-distributed locations results in overlapping ground station coverage and offers little latency benefit. We propose an algorithm to select locations that maximize ground station coverage and minimize transmission latency.

\item
For known event locations, placing ground stations nearby achieves zero transmission latency. We find that each ground station covers a 1500 km radius, so the transmission latency is zero as long as a station is positioned within 1500 km of the event location.
\end{packeditemize}
}

\section{Background and Motivation}
\label{sec:motivation}
We provide background on satellite event detection and highlight
key shortcomings of current systems in achieving end-to-end low-latency detection with low cost.

\subsection{Problem: low-latency event detection}
\label{ssec:problem}
%
%

\mypara{Nanosatellite constellations}
%
Advancements in nanosatellite technology and reductions in launch costs have
greatly increased the number of satellites in the low-earth orbit (LEO).  A
nanosatellite uses commercial off-the-shelf components (\eg
Planet~\cite{hsd2-planet-19-smallsat, planet-result-14-smallsat},
NASA~\cite{NASA-Qualcomm-Mars}) and has a small size and weight.  For example,
a "1U" CubeSat is $10cm \times 10cm \times 10cm$ and weighs a few
kilograms.  Nanosatellites are inexpensive enough to manufacture (a few
thousands dollars) and launch (tens of thousands of
dollars~\cite{leo-cubesat-cost}) in large numbers at a reasonable total cost.
Organizations such as Planet~\cite{Planet-homepage} and
Spire~\cite{Spire-homepage}, have deployed constellations of 
many nanosatellites with a common purpose.  These
constellations offer broad coverage and frequent revisits to the Earth's
surface compared to conventional, monolithic satellites.  Planet's
Dove Earth observation constellation has around 160 cubesats, providing
daily coverage of Earth, while NASA Landsat 9 (at \$750
million~\cite{landsat9-cost}) takes 16 days to image Earth.

\mypara{Orbit edge computing}
%
Advancements in machine learning and orbital edge computing have further unleashed the potential of Earth observation constellations, enabling them to process images directly on orbit.
For decades, Earth imaging satellites have downlinked all data to the ground 
 for processing. 
A satellite (and by extension a constellation) can collect far more data per orbit than it can downlink~\cite{kodan-asplos23}.  The downlink bottleneck prevents downlinking all images and a satellite must choose which a subset of images to downlink; a satellite may be forced to discard all but 10-20\% of images collected.
The need to subset data based on downlink constraints reduces a satellite's effective coverage area and limits its utility for Earth observation applications.
Prior work~\cite{oec-asplos20, sat-cal2019, kodan-asplos23} proposes to run computation tasks (\eg machine learning models for image classification and object detection) on satellites, to filter out low-value images and only send high-value images back to Earth, effectively alleviating the downlink bottleneck.

\mypara{Critical event detection}
Many work has shown that satellite images can be used for the detection of critical events, such as
forest fire~\cite{fire-detect-multiple-kernel, fire-sat-dataset}, 
oil spill~\cite{nasa-satellite-image-oil-spill, oil-spill-mdpi2019}, 
flood~\cite{sat-flood-detect-22-mdpi, sat-flood-detect-20-china-mdpi},
hurricane~\cite{cubesat-constellation-design-hurricane},
earthquake signature~\cite{2016-cubesat-earthquake-detect-possibility, 2002-cubesat-earthquake-detect-smallsat}
.
Previously, the image process pipeline was offline: images were downloaded from satellites to ground stations and subsequently processed with various machine learning techniques, which took hours or even days, posing challenges for timely responses.
Given the severe consequences of critical events on human lives and the environment, we consider whether we can leverage the new nanosatellite and orbit edge computing technology to provide low-latency event detection, to allow people more time to react.

\mypara{Components of Latency}
End-to-end latency is the duration of time from when an event first occurs until the ground reports to an operator that the event has occurred. 
As shown in \autoref{fig:design:sys_model}, the end-to-end latency
is composed of three components: capture latency, computation latency, and transmission latency.
Capture latency is the duration from the initiation of the event until a satellite passes over the region encompassing the critical event.
Computation latency is the time it takes for the ML model to run to detect the event.
Transmission latency is the time it takes the satellite to pass over a ground station and successfully transmit the image back to the ground.

We consider an existing Planet deployment: Dove~\cite{Planet-dove-constellation}, one of the largest cubesat earth observation constellations, with 160 satellites and 12 ground stations.
We use historical fire locations and simulate the constellation trajectory to calculate the end-to-end latency of fire detection (detail setup in \autoref{sec:impl}).
As shown in \autoref{fig:motivation:planet_baseline}, it takes around 7.5 (22) hours for the constellation to detect 50\% (95\%) fires.
In an intellectual contrast with prior work that has focused on optimizing transmission latency~\cite{l2d2-sigcomm, hotnet20-without-ISL, network-topology-design-conext19-eth, OrbitCast-icpn21-zeqi}, we find that the capture latency is the bottleneck: making up more than \naiveCaptureRatio\ of the end-to-end latency.
Compute latency is typically seconds or tens of seconds (details in
\autoref{ssec:eval:compute}), which is relatively small compared to capture and
transmission latency. Also, compute latency is
shorter than the frame arrival interval, and we do not consider queueing and
distribution of processing jobs, which has been studied in prior
work~\cite{oec-asplos20,kodan-asplos23} that is complementary.






\subsection{Prior approaches are insufficient}
\label{ssec:existing_solution}


%
To the best of our knowledge, no prior work has considered reducing capture latency.
Recent efforts~\cite{l2d2-sigcomm, deepak-umbra-mobicom23, deepak-nsdi24-serval} in nanosatellites focus solely on reducing transmission latency, which is less than 10\% of end-to-end latency. Consequently, eliminating transmission latency yields only marginal improvements.
These solutions also incur higher costs due to additional satellite or ground resources. For example, L2D2~\cite{l2d2-sigcomm} uses around 200 geo-distributed ground stations, increasing deployment and management expenses. OrbitCast~\cite{OrbitCast-icpn21-zeqi} relies on communication constellations like Starlink as relays for downloading data from observation satellites, adding rental costs.


Previous studies~\cite{walker-Constellations-1984, Rosette-Constellations-1980, yuanjie-hotnets21-cyber, leonet23-network-orbit-design} have explored the impact of constellation orbit design on communication satellites but have overlooked capture latency. In their scenarios, satellites must be within sight of the ground target to establish communication, rendering capture latency irrelevant (it is always zero).
Studies on observation satellites also overlook the impact of orbit design on capture latency. Prior work~\cite{cubesat-constellation-design-hurricane} considers orbit design to meet periodic regional revisit requirements (e.g., one visit every eight hours), but does not aim to reduce latency. Another study~\cite{smallsat-2021-monitor-india} reports capture latency for a single orbit design without exploring the effects of different orbit designs on capture latency.

\section{Design Space}

\begin{figure}[t]
    \centering
    \includegraphics[width=\linewidth]{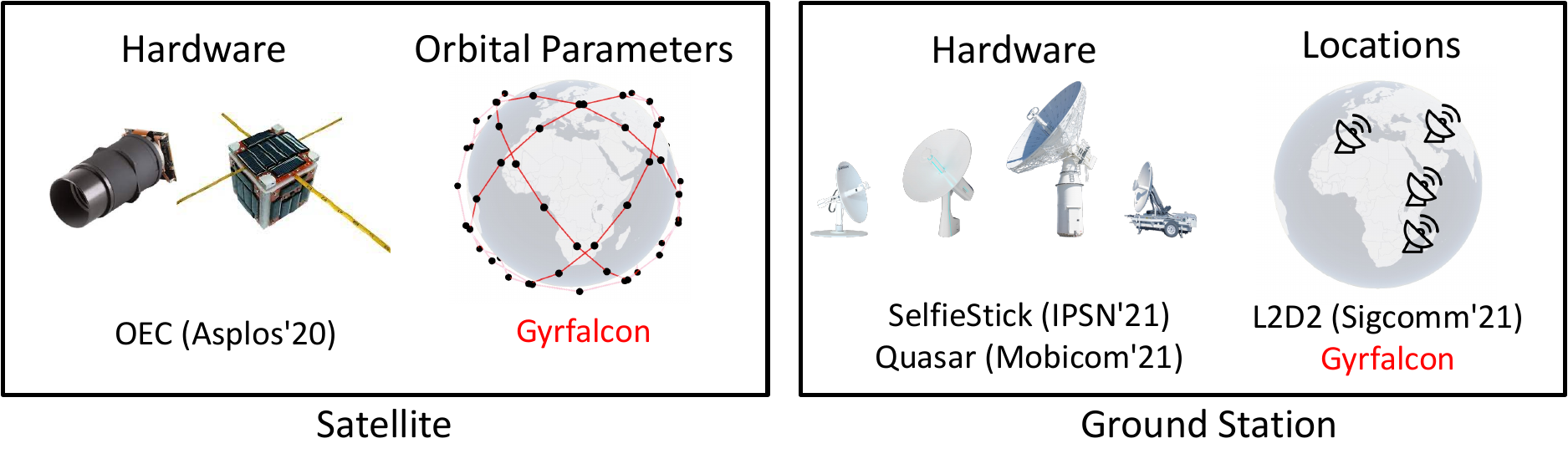}
    \caption{Nanosatellite constellation design space}
    \label{fig:design:design_space}
\end{figure}

As shown in \autoref{fig:design:design_space}, a nanosatellite constellation design contains two parts: satellite design and ground station design.

\mypara{Satellite Design}
A satellite design encompasses two key elements: satellite hardware selection and orbital parameter configuration.
A typical cubesat hardware suite comprises several components, including an Earth imaging sensor, a solar panel for energy harvesting, a capacitor or battery for energy storage, an attitude determination and control system for orientation adjustments, a transceiver for ground communication, and an onboard computer for control and orbital edge computing.

Hardware selection primarily hinges on mission objectives and budget constraints~\cite{oec-asplos20} and our approach makes relatively few assumptions about the hardware design of nanosatellites. 
For example, agricultural monitoring may favor a high-resolution visual spectrum sensor, while forest fire monitoring may require a low-resolution, wide-coverage, short-wave infrared sensor.
The sensor choice can be made based on the prior image data~\cite{oec-asplos20, sat-cal2019, eagleeye-asplos24}.
Another example is that deployable solar panels, though more costly, generate more energy than mountable panels.
A depleted energy supply in the cubesat would hinder its ability to capture, process, or transmit images.
We assume sufficient energy from solar panels for continuous operation, treating energy as a separate issue. Other research addresses energy concerns by dividing workloads among satellites~\cite{oec-asplos20} or using large, deployable solar panels~\cite{Planet-dove-constellation}.

The orbital parameters, such as inclination, number of planes, plane distribution, decide which regions are covered by the constellation and how often a region is revisited, which is directly related to capture latency.
Also, nanosatellites often lack sufficient fuel reserves because of their small size and hence hardly change orbits.
\ul{
Therefore, we aim to choose the best orbital configuration for launch.
}

\mypara{Ground Station Design}
Ground station design encompasses both hardware selection and location determination.
Hardware selection includes critical factors such as frequency band, power and dish size. These choices are also influenced by the mission's objectives and budget constraints~\cite{Vaibhav-mobicom21-Quasar,metasurface-satellite-hotnet22,liliqiu-metasurface-mobicom23}.
For example, the frequency band should align with the satellite transceiver frequency band. 
A larger, higher-power dish, while more expensive, can offer increased signal strength, leading to higher communication bandwidth.

Ground station locations decide when a satellite can communicate with ground because a satellite's elevation degree must be greater than a threshold (\eg $10^{\circ}$) for communication to occur. 
Therefore, ground station locations would affect transmission latency.
Existing commercial deployments choose to put ground stations near the polar~\cite{Planet-homepage, ksat-ground-station}
to allow satellites to communicate with a ground station during every period.


Also, ground station location options are limited to regions with reliable Internet connectivity, which is essential to transmit satellite signals back to the operators. This constraint prevents the deployment of ground stations in remote areas lacking sufficient population density and Internet infrastructure, such as oceans or mountains. 
In addition to building their own ground stations, operators could also choose to rent ground station service from providers such as AWS Ground Station~\cite{amazon-ground-station}.
These providers handle the construction and management of ground station infrastructure, yet customers remain responsible for determining the appropriate times and locations for renting ground stations.
\ul{
We consider how to choose the best locations based on a large set of feasible locations.
}

\mybox{
Takeaway: Satellite hardware and ground station hardware are selected based on mission goals and budget constraints. In this paper, we focus on designing satellite orbit parameters and ground station locations.
We present a measurement study on how these factors influence latency, offering guidance (we name it as \SYS) to operators on optimal design.
}


\section{Measurement Methodology}
\label{sec:impl}

In this section, we detail our simulation-based measurement methods and the datasets from several real-world use cases.

\subsection{Simulation Method}

We use the Cote simulator, which has also been used by prior work~\cite{oec-asplos20, kodan-asplos23, eagleeye-asplos24}, to compute the latency.
The simulator model physical-based SGP4 orbits~\cite{sgp4-model} for satellites using orbit
parameters to generate satellite locations over time. 
By comparing satellite, event, and ground station locations, we can determine whether a satellite can capture an event or communicate with a ground station.
We employ a Monte Carlo simulation to calculate the capture and transmission latency for each constellation and ground station configuration.
For each run, the simulator randomly selects events locations based on historical
locations and assigns a start time to each event.  We assume that an event will
not disappear once it starts because events like fires, landslides usually last days.  

\change{
We use simulation instead of real deployment due to the high cost (hundreds of millions of dollars) of deploying a constellation with over 100 satellites and 10 ground stations. Our simulation is based on a physical model, providing accurate latency results. Simulating satellite trajectories is a common standard in the satellite industry prior to real deployment~\cite{oec-asplos20, eagleeye-asplos24, l2d2-sigcomm, deepak-nsdi24-serval, deepak-umbra-mobicom23}. We believe that the lessons learned from the simulation will provide valuable guidance for future industry deployments.
}

%


We consider satellites with identical cameras with 125km swath,
although \SYS also applies to heterogeneous constellations.
We consider an existing Planet constellation with 160 satellites and 12 Planet polar ground stations.
We obtain orbit parameters for Planet satellites from
Celestrak~\cite{celestrak-tles}.
We use wildfire images from prior work~\cite{fire-sat-dataset} captured by
Landsat 8, wht 30m GSD. We train several segmentation models~\cite{segmentation_models-pytorch-github} for fire detection tasks.  
We consider a frame size of 4096 $\times$ 4096, tiled to 256 $\times$ 256,
processing one tile at a time. 
For ground station, we consider 12 Planet ground stations and 264 SatNOGS~\cite{satnogs-ground-station} ground stations used by L2D2~\cite{l2d2-sigcomm}.

\subsection{Evaluation Use Cases}
\label{ssec:use_case}

\begin{figure}[t]
    \centering
    \includegraphics[width=0.97\linewidth]{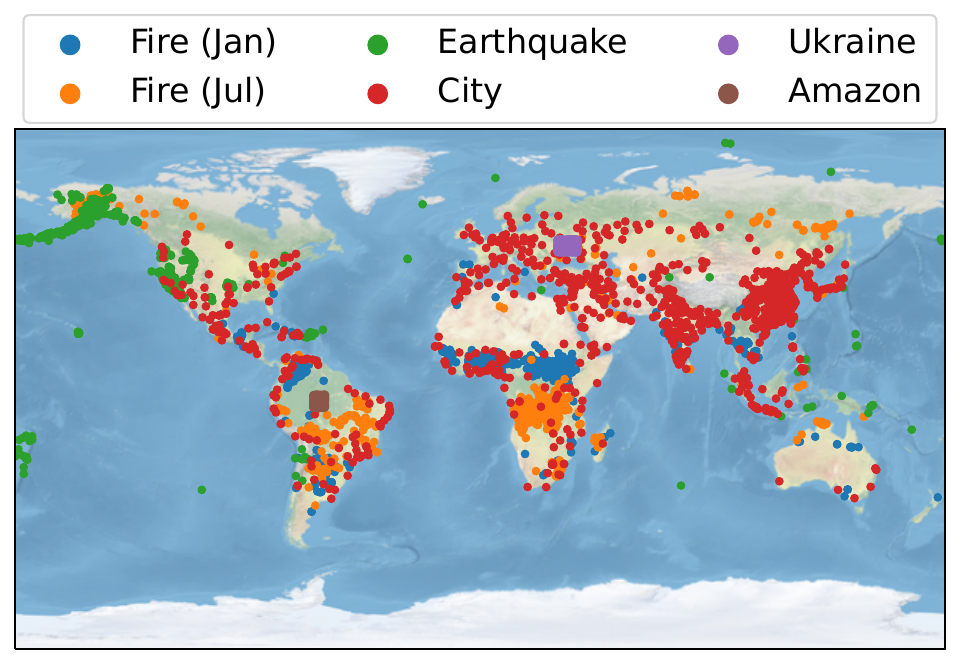}
    \caption{We evaluate \SYS with six use cases with varying location distributions.}
    \label{fig:eval:visual_multi_event}
\end{figure}

We consider several event use cases and
\autoref{fig:eval:visual_multi_event} shows the event locations.
{\noindent \bf Fire Detection.} We consider an application that uses satellites
to detect forest fires.  We obtain active fire locations from
NASA~\cite{nasa-fire-location}, which contains \mytextapprox 8 million fires
that occurred in 2021-2022. 
Since fire locations vary with the seasons, we consider two different scenarios: fires in January and fires in July.
{\noindent \bf Earthquake Signature Detection.}
We consider an application that uses satellites to detect earthquake signature.
We obtain historical earthquake locations from the
USGS~\cite{usgs-earthquakes-map}.
{\noindent \bf City Monitoring.}
We consider an application that uses satellites to monitor cities for traffic
patterns and infrastructure integrity.  We obtain the locations of the top 1000
most populous cities from GeoNames~\cite{city-population-dataset}.
{\noindent \bf Geo-clustered event tracking.}
We created two synthetic, geographically clustered datasets.  In the first
dataset, all events occur uniformly randomly within a \SI{500}{km} $\times$
\SI{500}{km} region of Kyiv (Ukraine), representative of a use case related to
wildlife management and tracking in eastern Ukraine.  In the second dataset,
all events occur uniformly randomly within a \SI{500}{km} $\times$ \SI{500}{km}
region of the Amazon rainforest in Brazil, representative of a use case related
to forestry management and identifying illegal deforestation.  
{\noindent \bf Uniform Grid.}
We consider events to be uniformly distributed worldwide, representing a use case where events are not concentrated in any specific location.
We divide the globe into a uniform grid~\cite{revisit-interval-sentinel-2017} and consider each cell central point as an event location.

%
%


%
%

\section{Reducing Capture Latency}
\label{sec:capture_latency}

We first show some constellation examples. Then, we list the orbital parameters and provide a detailed measurement study on how the parameters affect the capture latency. Finally, we discuss the potential constraints on using different parameters.

Previous work~\cite{leonet23-network-orbit-design} has shown how
constellation orbit design affects communication satellites (\eg Starlink), but has overlooked capture latency. For communication, satellites must be within sight of the ground target to establish
communication, rendering capture latency irrelevant (it is
always zero). In contrast, for earth observation tasks, the capture latency could be nonzero (\eg one satellite flies over a forest fire region after the fire starts).

\begin{figure}
\begin{subfigure}[t]{0.22\linewidth}
   \centering
   \includegraphics[width=0.9\linewidth]{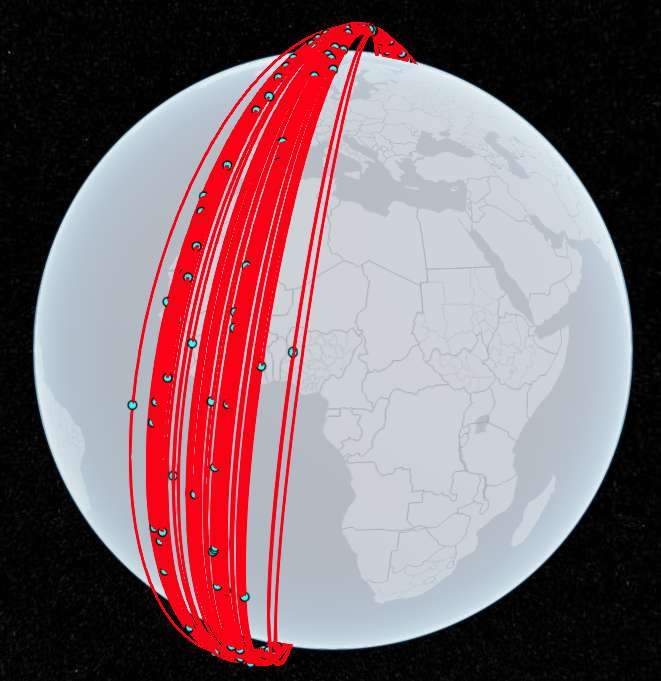}
   \caption{Planet}
   \label{fig:orbit:planet}
\end{subfigure}
\begin{subfigure}[t]{0.22\linewidth}
   \centering
   \includegraphics[width=0.9\linewidth]{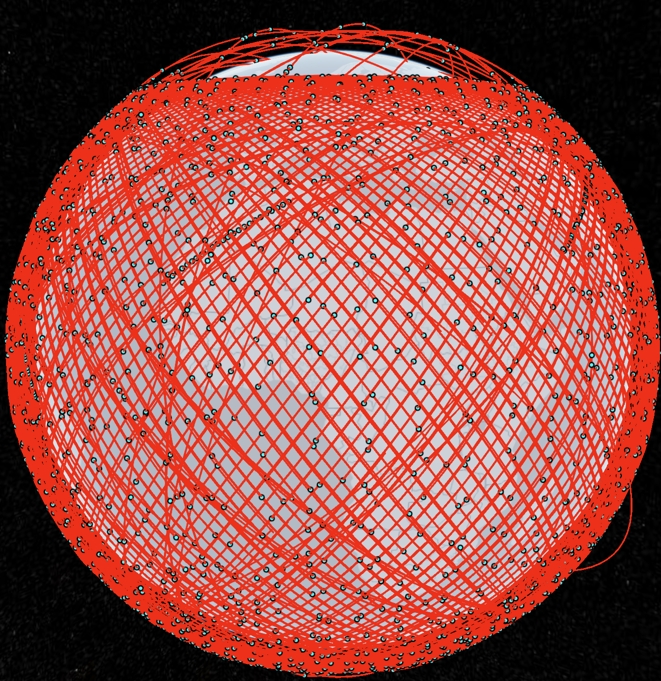}
   \caption{Starlink}
   \label{fig:orbit:starlink}
\end{subfigure}
\begin{subfigure}[t]{0.22\linewidth}
   \centering
   \includegraphics[width=0.9\linewidth]{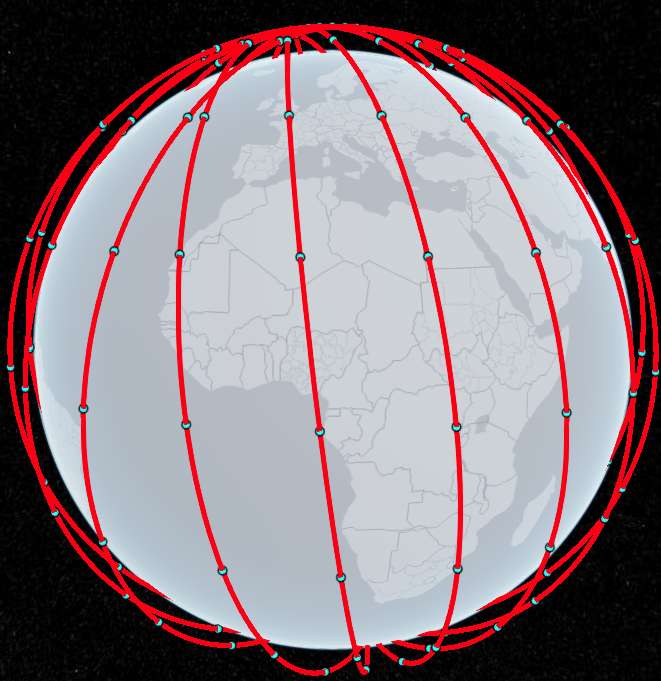}
   \caption{\SYS (Fresh)}
   \label{fig:orbit:fresh}
\end{subfigure}
\begin{subfigure}[t]{0.22\linewidth}
   \centering
   \includegraphics[width=0.9\linewidth]{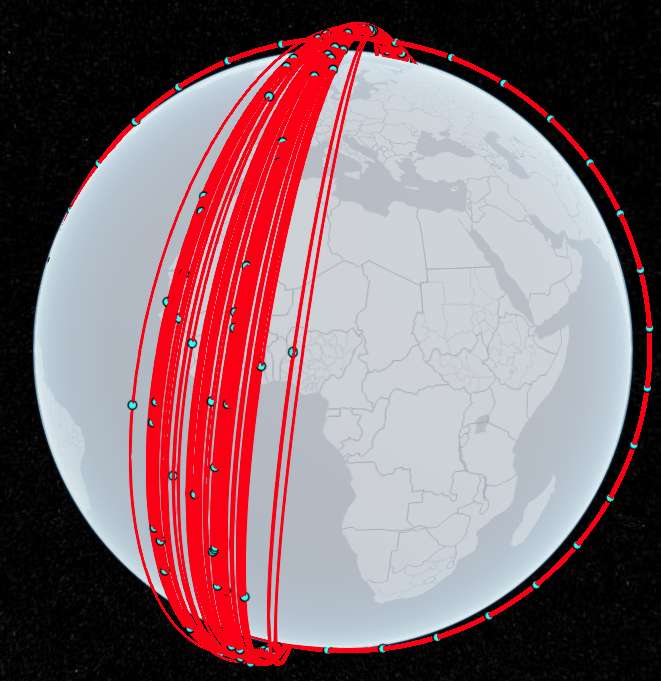}
   \caption{\SYS (Incremental)}
   \label{fig:orbit:incremental}
\end{subfigure}

\centering
\caption{
Orbital configuration visualization across constellations.
}
\label{fig:eval:orbit_visual}
\end{figure}

\subsection{Constellation Orbital Configuration Examples}

\autoref{fig:eval:orbit_visual} graphically compares the orbits of existing
constellations and those designed by \SYS. 
The Planet earth-observation constellation with 160 satellites follows almost the same plane, which leaves many regions unmonitored for a long time.
\change{
The current Planet constellation design does not optimize for event detection latency because their business model focuses on providing daily Earth images, and their image processing pipeline takes hours to days before an image reaches the end user~\cite{deepak-umbra-mobicom23, deepak-nsdi24-serval}. Given the opportunity that edge computing allows real time event detection on the satellite, we consider how to design optimal orbital configuration to reduce capture latency.
}

%
On the other hand, the Starlink communication constellation with \mytextapprox 6000 satellites both deploys satellites into different orbits to
maximize coverage.
We also show the fresh and incremental deployment constellations with 160 satellites, selected under the \SYS guideline to minimize capture latency, in \autoref{fig:eval:orbit_visual}(c)(d).

\begin{figure*}[t]
    \begin{minipage}{0.66\linewidth}
    \begin{subfigure}[t]{0.23\linewidth}
        \centering
        \includegraphics[width=0.9\linewidth]{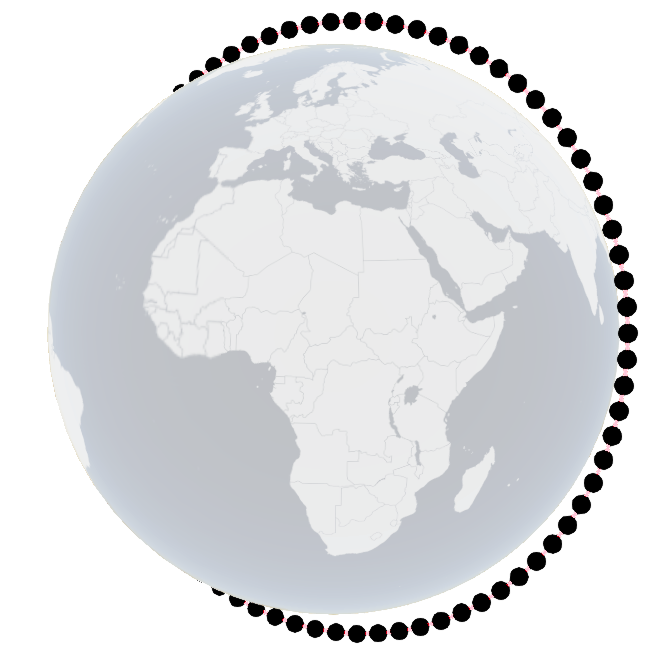}
        \caption{1 Plane ($97^{\circ}$)}
        \label{fig:orbit:plane1}
    \end{subfigure}
    \begin{subfigure}[t]{0.23\linewidth}
        \centering
        \includegraphics[width=0.9\linewidth]{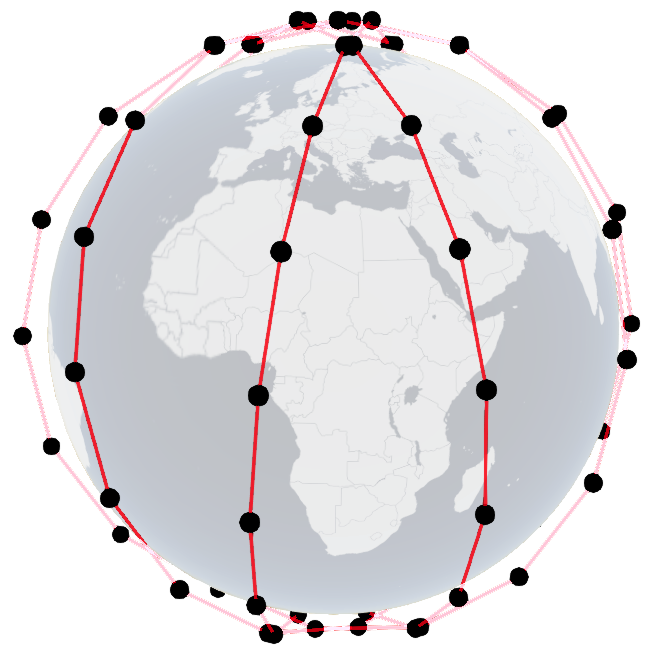}
        \caption{5 Plane ($97^{\circ}$)}
        \label{fig:orbit:plane5}
    \end{subfigure}
    \begin{subfigure}[t]{0.23\linewidth}
        \centering
        \includegraphics[width=0.9\linewidth]{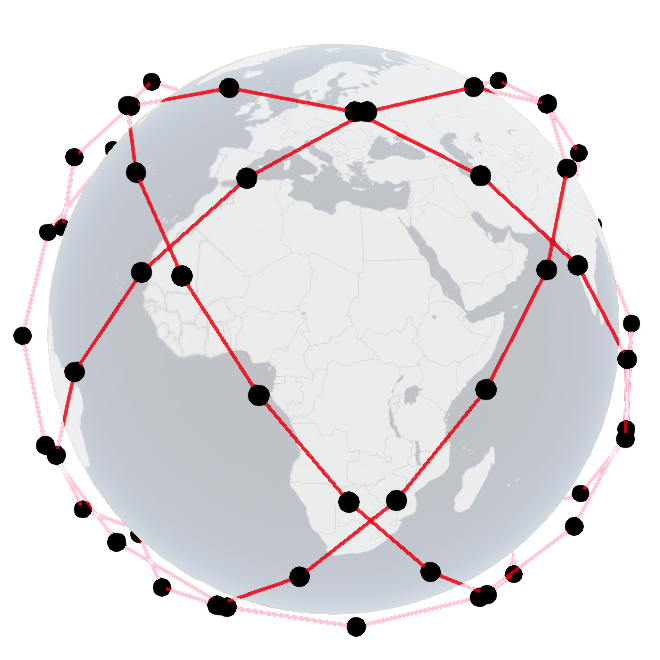}
        \caption{5 Plane ($53^{\circ}$)}
        \label{fig:orbit:plane5_53}
    \end{subfigure}
    \begin{subfigure}[t]{0.23\linewidth}
        \centering
        \includegraphics[width=0.9\linewidth]{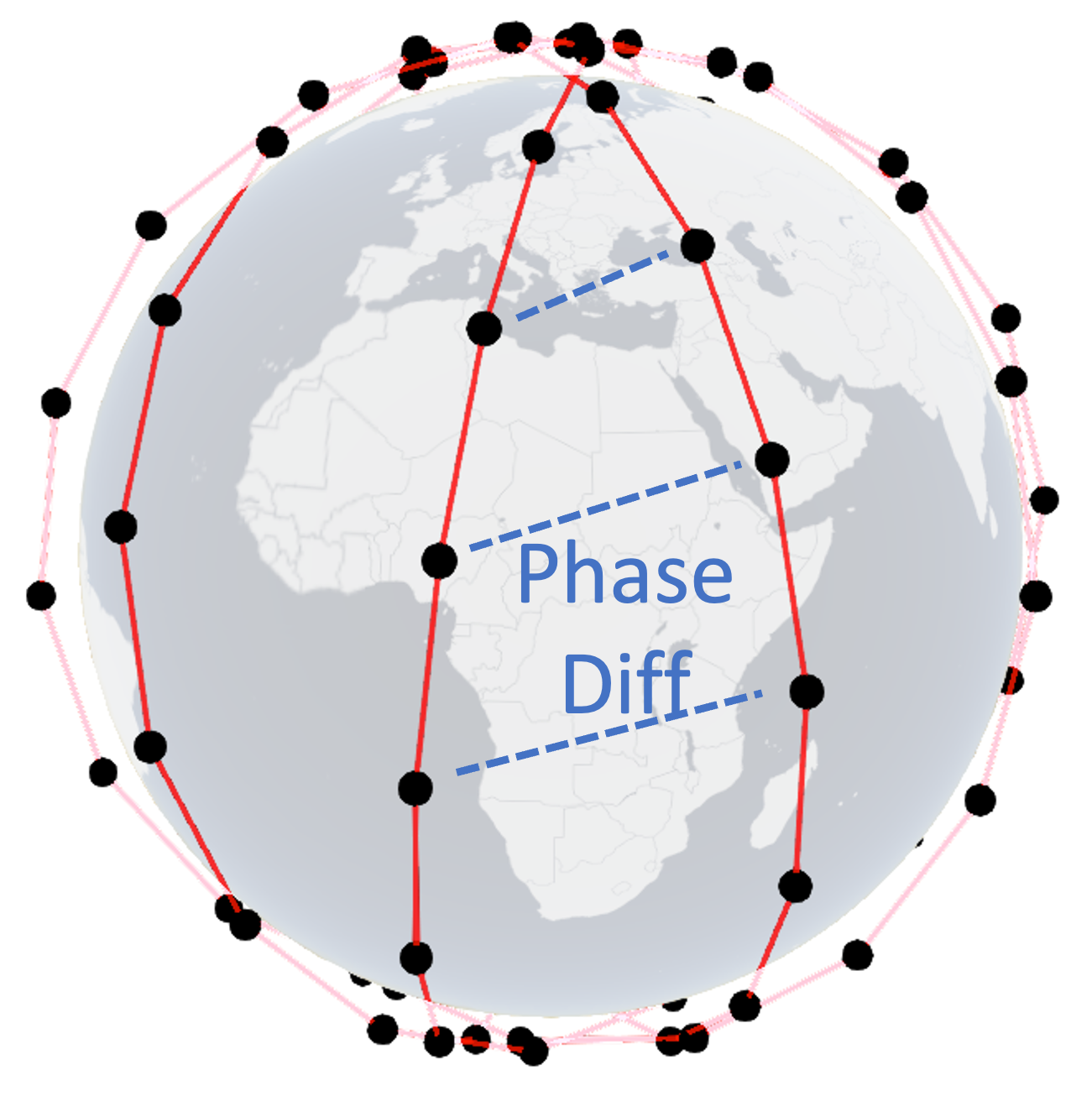}
        \caption{5 Plane($97^{\circ}$ w/ phase diff)}
        \label{fig:orbit:plane5_phase}
    \end{subfigure}
    \caption{
    Example orbital configurations for a constellation.
    }
    \label{fig:orbit_visual}
    \end{minipage}
    \begin{minipage}{0.33\linewidth}
    \begin{subfigure}[b]{0.48\linewidth}
        \centering
        \includegraphics[width=0.9\linewidth]{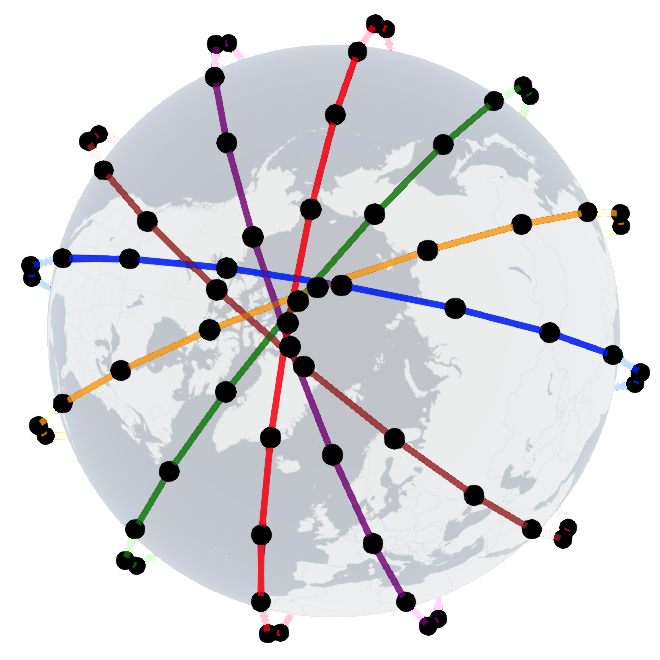}
    \end{subfigure}
    \begin{subfigure}[b]{0.48\linewidth}
        \centering
        \includegraphics[width=0.9\linewidth]{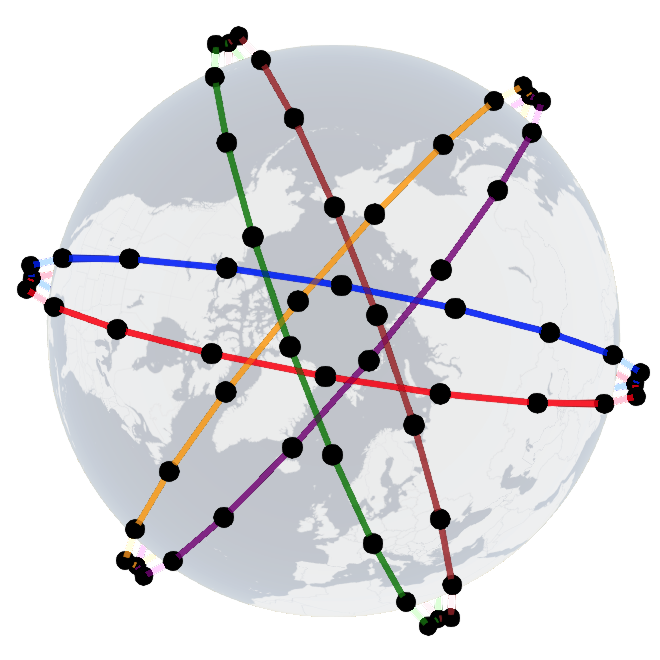}
    \end{subfigure}
    \caption{Example configurations for plane distribution (north polar view): split planes evenly across 180\mydeg (Left) or 360\mydeg (Right). 
    }
    \label{fig:orbit_visual_split_plane}
    \end{minipage}
\end{figure*}

\subsection{Satellite Orbit Parameters}
\label{ssec:orbital_para}

\begin{table}[t]
\footnotesize
\centering
\begin{tabular}{|p{0.24\linewidth} | p{0.69\linewidth}|}
\hline

Parameter  &  Description  \\ 
\specialrule{0.2em}{0.1em}{0.1em}
Inclination  & Angle between the orbit and the Earth equator plane.  \\ \hline
Altitude  & Satellite height above Earth.  \\ \hline
Eccentricity  & A measure of how elliptical the orbit is.  \\ \hline
Number of plane  &  Quantity of orbital planes. \\ \hline
Plane phase  &  Phase difference between planes.  \\ \hline
Plane distribution & 
Manner in which orbital planes are distributed.
\\ \hline
\end{tabular}
    \caption{Summary of Orbit Parameters}
    \label{table:orbit_parameters}
\end{table}

We list the parameters that determine orbit configurations (summarized in \autoref{table:orbit_parameters}) and analyze their impact on capture latency.
\autoref{fig:orbit_visual} and \autoref{fig:orbit_visual_split_plane}
visualize some example orbital configurations with different parameters.
Overall, we find that the number of planes is the most important factor for capture latency.

\begin{figure}[t]
    \centering
    \includegraphics[width=0.95\linewidth]{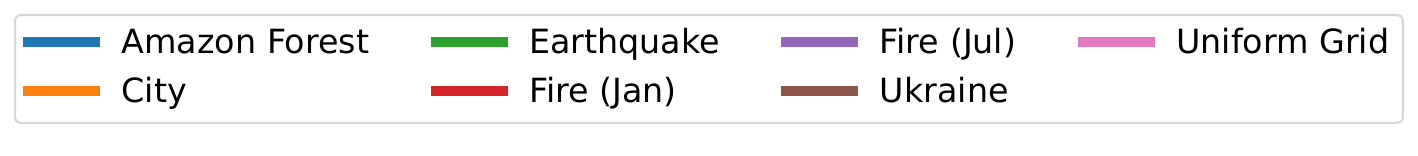}
    \includegraphics[width=0.95\linewidth]{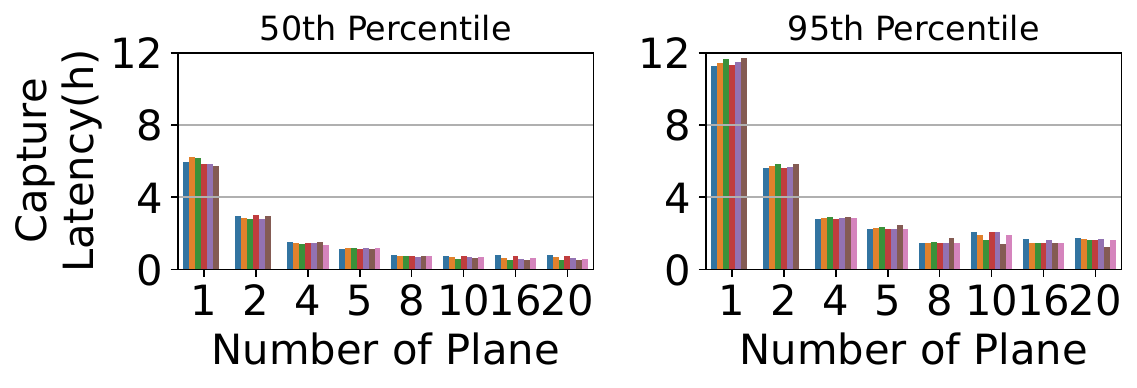}
    \caption{
    \textbf{Capture latency with different number of planes.}
    Using more planes decreases capture latency by \evalMorePlaneImprove, plateauing when there are more than 10 planes.}
    \label{fig:capture:plane_line_plot}
\end{figure}

\begin{figure}[t]
    \centering
    \includegraphics[width=0.95\linewidth]{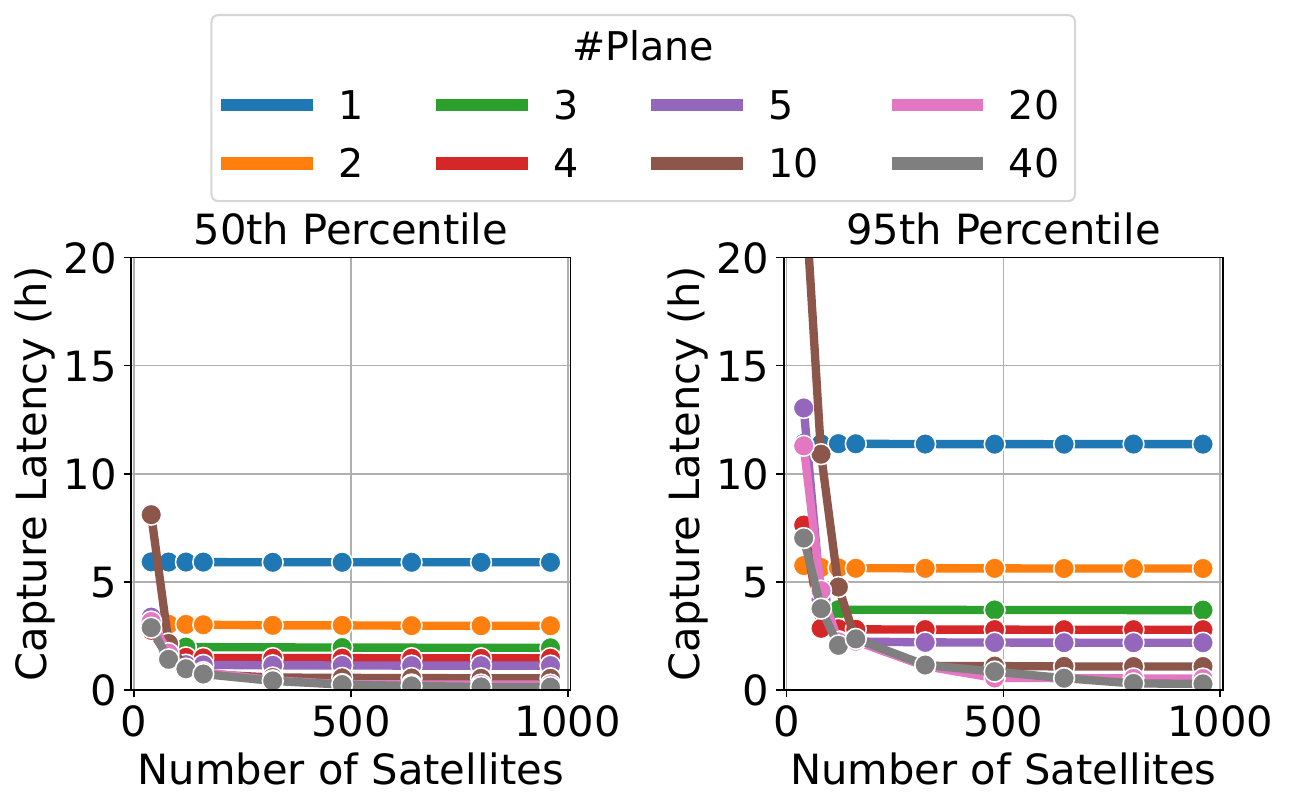}
    \caption{
    \textbf{Capture latency with different number of satellites and different number of planes.}
    More satellites reduce capture latency but reach a plateau. More planes reduce the plateau.
    }
    \label{fig:capture:plane_with_num_sat}
\end{figure}

\mypara{Number of planes}
\autoref{fig:orbit:plane1} and \autoref{fig:orbit:plane5} illustrate constellations with 1 plane and 5 planes.
Deploying constellations in multiple planes increases coverage.
For example, the Starlink constellation uses 72 planes.

\autoref{fig:capture:plane_line_plot} shows the capture latency when using different numbers of planes.
Using 10 planes reduces the latency by \evalMorePlaneImprove\ compared to using 1 plane, because it reduces the overlap between satellites from different planes.
It also shows that using more than 10 planes does not further reduce the latency.
\autoref{fig:capture:plane_with_num_sat} shows how the number of satellites affects the plateau.
While increasing the number of satellites initially reduces capture latency, it eventually plateaus due to overlapping capture regions. 
Incorporating more planes mitigates this plateau effect.

\begin{figure}[t]
    \centering
    \includegraphics[width=0.66\linewidth]{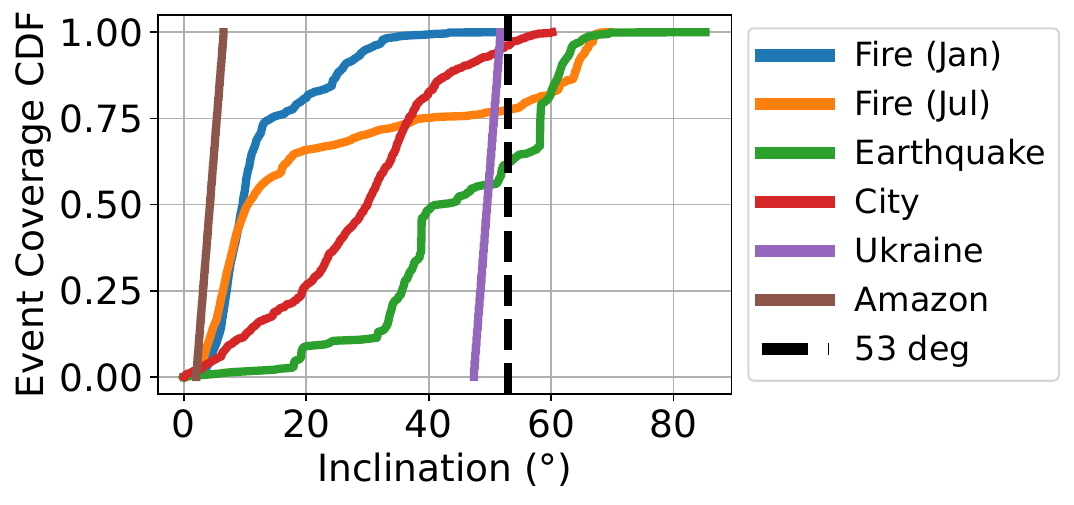}
    \caption{ \label{fig:capture:inclination_coverage}
    \textbf{Event coverage with different inclination.}
    Lower inclination cannot cover all events.
    }
\end{figure}

\begin{figure}[t]
    \begin{subfigure}{0.95\linewidth}
        \centering
        \includegraphics[width=0.95\linewidth]{fig/eval/capture_latency_legend.pdf}
        \includegraphics[width=\linewidth]{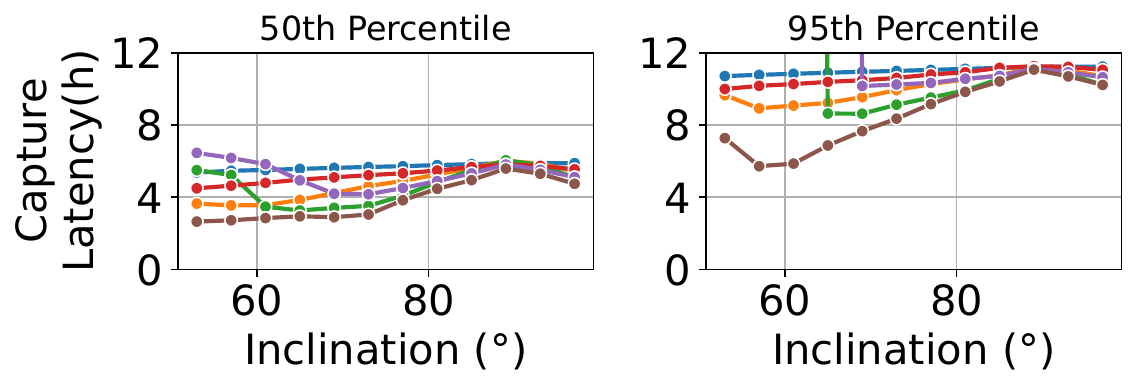}
        \caption{With 2 Planes}
    \end{subfigure}
    \begin{subfigure}{0.95\linewidth}
        \centering
        \includegraphics[width=\linewidth]{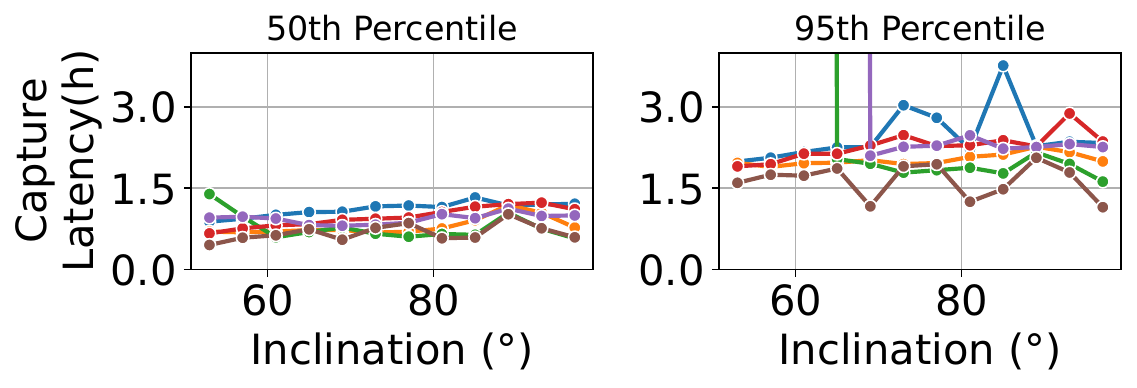}
        \caption{With 10 Planes}
    \end{subfigure}
    \caption{
    \textbf{Capture latency with different number of planes.}
    Lower inclination reduces capture latency but may miss some events. The reduction diminishes as more planes are employed.
    }
    \label{fig:capture:change_inclination}
\end{figure}

\mypara{Inclination}
Inclination is the angle between a satellite's orbit and Earth's equatorial plane.
\autoref{fig:orbit:plane5} and \autoref{fig:orbit:plane5_53} illustrate
constellations at different inclinations.
Inclination affects coverage.
The Planet constellation uses 
97\mydeg\ inclination, providing global coverage with the exception of a small
polar region.  
The Starlink constellation uses a mix inclination figuration (53\mydeg and 97\mydeg) where the 97\mydeg orbit provides global coverage and 
the 53\mydeg\ orbits spend more time and provide better services for middle latitudes where most customers live.
\autoref{fig:capture:inclination_coverage} shows the event coverage using constellations with different inclinations. 
Using a lower inclination reduces coverage.
For example, using 53$^{\circ}$ inclination miss 22\%, 38\% events for two use cases.

\autoref{fig:capture:change_inclination} shows that a lower inclination can reduce the capture latency when using a small number (\ie 2) of planes.
But it also results in a high 95th percentile latency for some use cases because the low-inclination constellation cannot cover some events in these use cases, resulting in an infinite capture latency for those events.
On the other hand, when using a large number (\ie 10) of planes, the benefit with a lower inclination diminish.
Due to the page limit, we only show using 2 and 10 planes here, but we observe a similar trend for other numbers of planes.

\begin{figure}[t]
    \centering
    \includegraphics[width=0.95\linewidth]{fig/eval/capture_latency_legend.pdf}
    \includegraphics[width=0.97\linewidth]{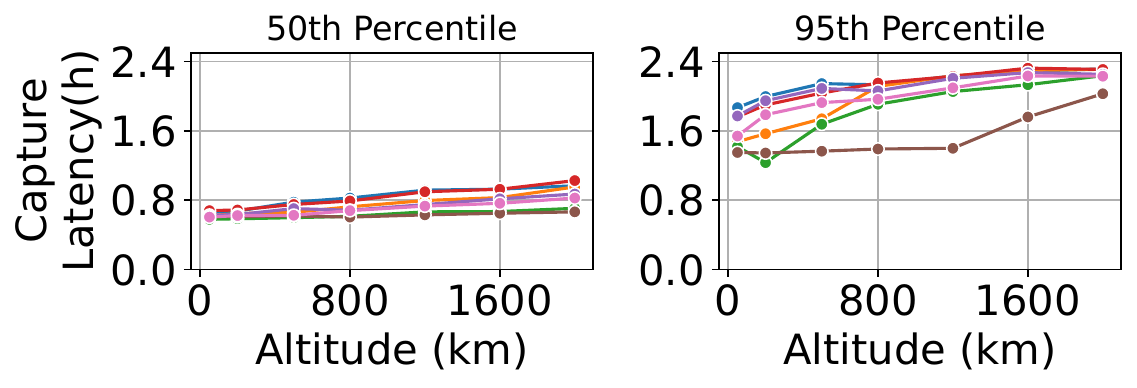}
    \caption{
    Lower altitudes reduce capture latency.
    }
    \label{fig:eval:capture_latency_altitude}
\end{figure}

\change{
\mypara{Altitude}
Altitude is a satellite's orbit height above Earth's surface.
Low earth orbit (LEO) satellites typically have a altitude ranging from 200km to 1000km. 
\autoref{fig:eval:capture_latency_altitude} shows that lower altitudes reduce capture latency.
This is because satellites at lower altitudes have shorter orbital periods (higher angular velocity) and can cover more areas in the same amount of time. For example, changing altitude from 200km to 1000km changes the period from 88min to 105min (1.2x).
However, altitude also affects the mission's lifetime. Most satellites operate around 500 km, yielding a 10-year lifetime. A lower altitude (e.g., 200 km) reduces the lifespan to 1 day~\cite{Orbital-Lifetimes} due to the increase in atmospheric drag. Extending the lifetime at a lower altitude requires costly failure-prone propulsion systems with additional certifications.
For the rest of the evaluation, we assume the satellites are at 500 km.
}


\begin{figure}[t]
    \centering
    \includegraphics[width=0.6\linewidth]{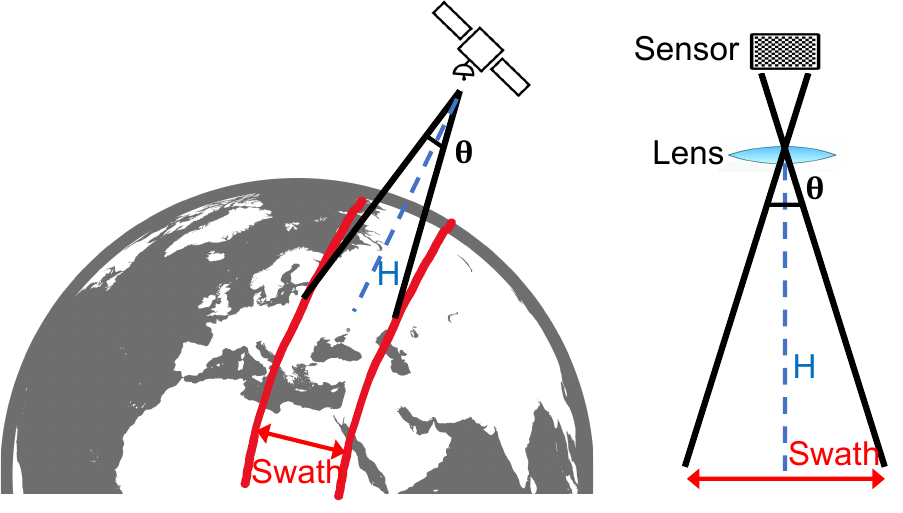}
    \caption{
    Swath is determined by altitude and camera field of view.
    }
    \label{fig:design:swath_illustrate}
\end{figure}

\begin{figure}[t]
    \centering
    \includegraphics[width=0.6\linewidth]{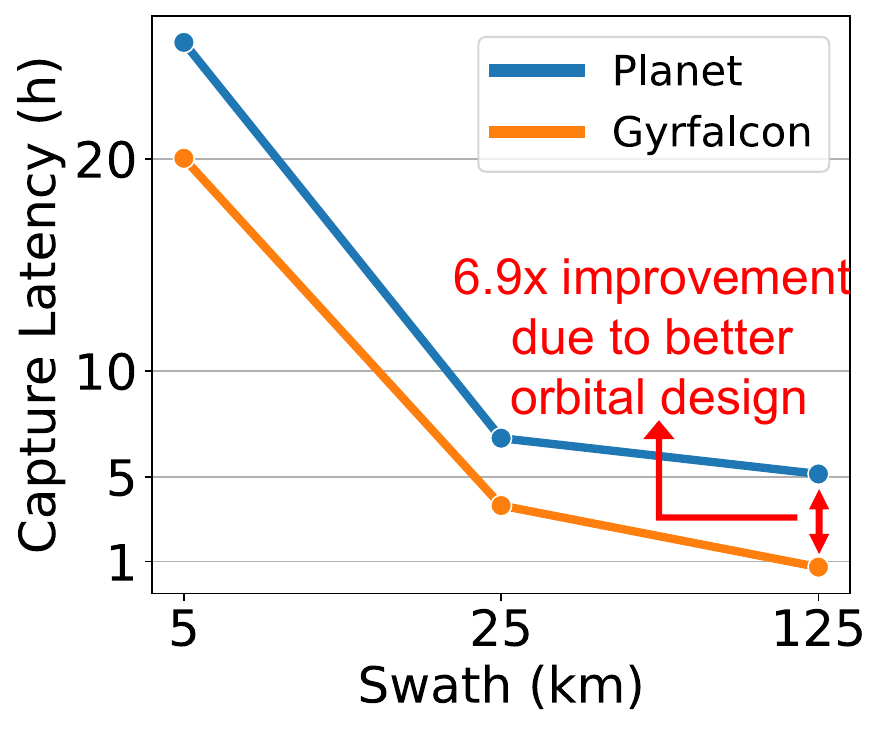}
    \caption{
    Larger swath reduce capture latency. \SYS improves orbital design, further reducing capture latency.
    }
    \label{fig:capture:planet_large_swath}
\end{figure}

\change{
\mypara{Discussion on Swath}
Swath is the horizontal distance covered by a satellite sensor as it captures images of the Earth's surface. A larger swath increases satellite coverage and reduces capture latency. However, swath is constrained by hardware limitations and application requirements.

The swath is related to the ground sample distance (GSD), which is the distance on the ground between two consecutive pixels in the satellite image, and the number of pixels in one direction of the sensor:
$$
\textit{Swath (km)} = \textit{GSD (km/px)} \times \textit{Pixel Count in One Side}
$$
Prior work~\cite{eagleeye-asplos24} shows that most commercial sensors have a fixed number of pixels in one side (\eg 4096), which means that a larger swath would result in a lower resolution. Combining multiple sensors could increase the pixel count, but would result in a camera with a large weight and volume, leading to a larger satellite with high launching cost.
Operators choose camera resolution based on prior image data and application requirements~\cite{oec-asplos20, sat-cal2019, eagleeye-asplos24}, which then leads to a fixed swath.
\autoref{fig:capture:planet_large_swath} shows how median capture latency
changes with different swaths. 
It shows that a larger swath reduces the capture latency and \SYS improves orbital design, further reducing capture latency.

On the other hand, there is a relationship between swath, altitude $H$, and camera field of view $\theta$ as shown in \autoref{fig:design:swath_illustrate}:
$$
\textit{swath} = H \times \tan\left(\frac{\theta}{2}\right)
$$
Simply using a higher altitude or a camera with a larger field of view to get a larger swath can lead to lower resolution. 
A better approach is for operators to choose the altitude based on the planned life of the satellite and federal regulations. Then, based on the required resolution (and consequently the swath), operators can then determine the optimal camera field of view.
}


%
\mypara{Eccentricity}
Eccentricity quantifies the extent to which an orbit deviates from a perfect circle. A satellite in an elliptical orbit flies at varying altitudes and speeds at different points in the ellipse. Nanosatellites are typically equipped with cameras that have a fixed field of view, as moving parts are prone to mission failure. This results in varying swath and GSD values when the satellite flies over different locations, with larger GSD values potentially rendering images unusable for some applications. Our exclusive focus is on circular orbits to ensure unbiased global coverage. A future direction is to explore elliptical orbits tailored to specific event locations.


\begin{figure}[t]
    \centering
    \begin{subfigure}{0.95\linewidth}
        \centering
        \includegraphics[width=0.95\linewidth]{fig/eval/capture_latency_legend.pdf}
        \includegraphics[width=\linewidth]{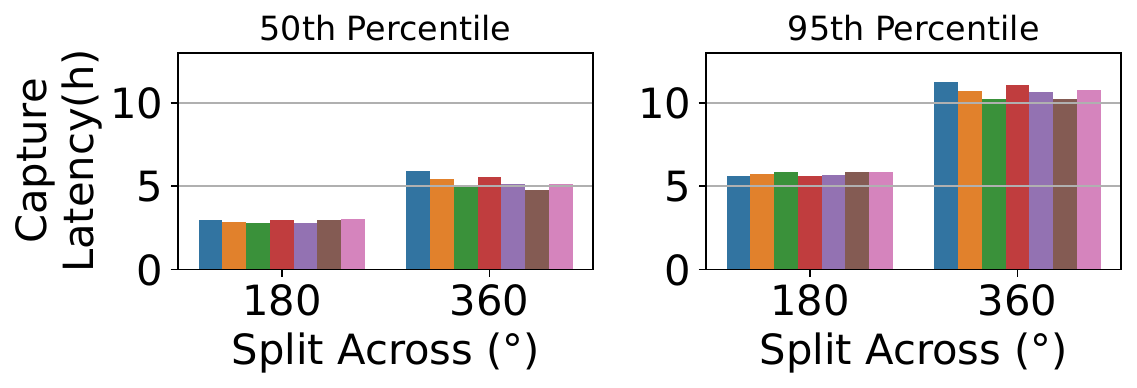}
        \caption{With 2 Planes}
    \end{subfigure}
    \begin{subfigure}{0.95\linewidth}
        \centering
        \includegraphics[width=\linewidth]{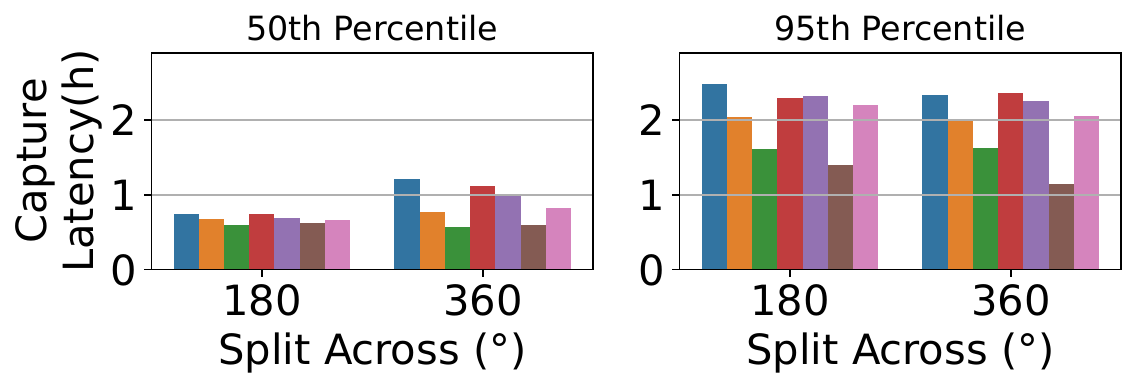}
        \caption{With 10 Planes}
    \end{subfigure}
    \caption{
    \textbf{Capture latency with different plane distribution.}
    Splitting planes across 180 \mydeg reduces capture latency.
    }
    \label{fig:capture:capture_plane_across}
\end{figure}

\begin{figure}[t]
    \centering
    \begin{subfigure}{0.95\linewidth}
        \centering
        \includegraphics[width=0.95\linewidth]{fig/eval/capture_latency_legend.pdf}
        \includegraphics[width=\linewidth]{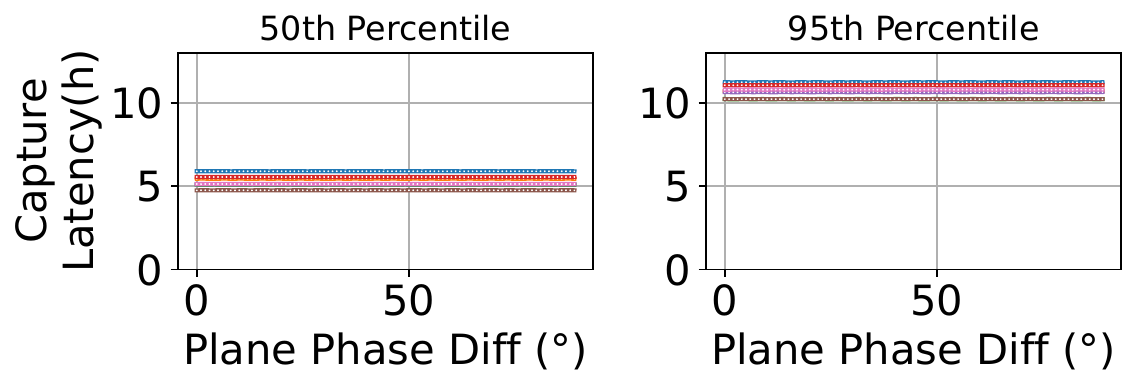}
        \caption{With 2 Planes}
    \end{subfigure}
    \begin{subfigure}{0.95\linewidth}
        \centering
        \includegraphics[width=\linewidth]{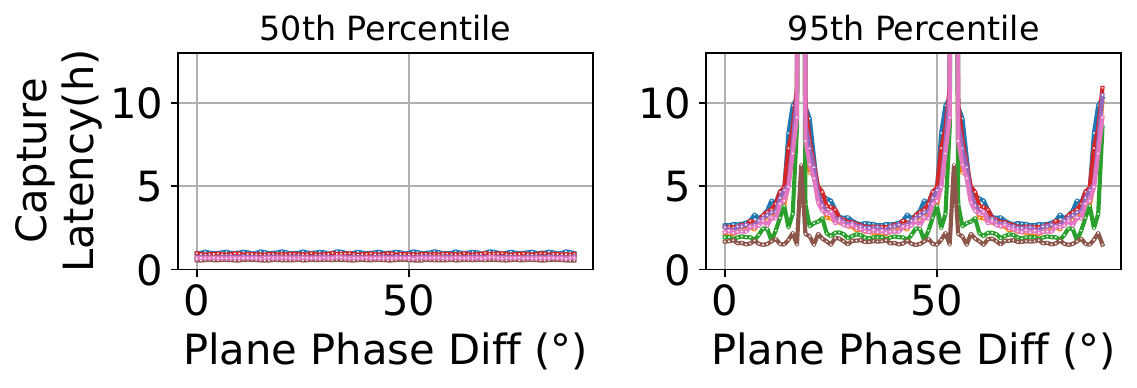}
        \caption{With 16 Planes}
    \end{subfigure}
    \caption{
    \textbf{Capture latency with different plane phase difference.}
    Plane phase difference impacts tail latency only in cases involving a large number of planes.
    }
    \label{fig:capture:capture_phase_diff}
\end{figure}

\mypara{Plane Phase}
The phase difference between planes (\autoref{fig:orbit:plane5_phase}) affects the frequency of overlap in the satellite coverage between planes.
We consider rotating each plane with a fixed degree of phase difference compared to the previous plane, similar to prior work~\cite{walker-Constellations-1984}.
\autoref{fig:capture:capture_phase_diff} shows that the plane phase does not affect the median capture latency and only affects the 95th latency when a large number of planes are used.
This is because more planes lead to closer proximity between planes, making the phase difference more important.

\mypara{Plane Distribution}
Prior work~\cite{walker-Constellations-1984} considers to distribute planes uniformly across 
180\mydeg  or 360\mydeg, as shown in \autoref{fig:orbit_visual_split_plane}.
Dividing planes across 360\mydeg may increase overlap between planes, whereas dividing them across 180\mydeg mitigates overlap by introducing some asymmetry.
\autoref{fig:capture:capture_plane_across} shows that splitting planes across 180$^{\circ}$ reduces the capture latency when using a small number of planes, but the difference is no longer significant when using many planes.

\begin{figure}[t]
    \centering
    \includegraphics[width=0.6\linewidth]{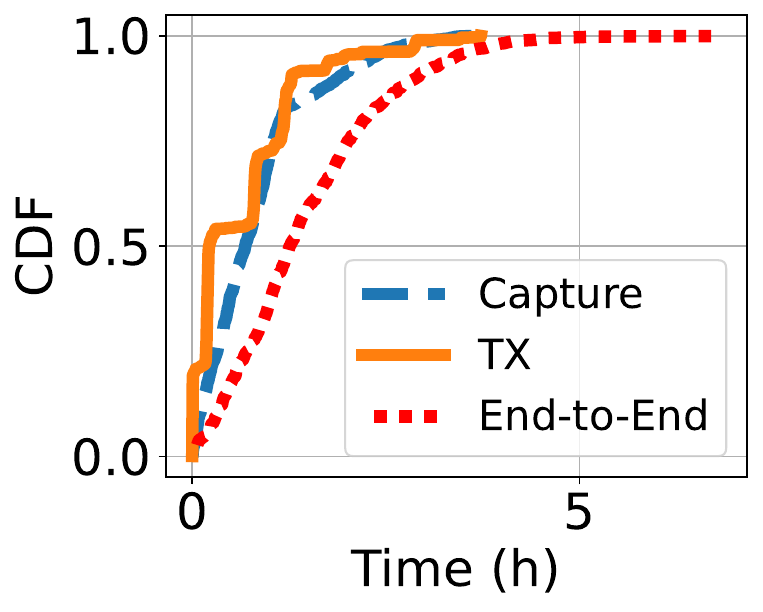}
    \caption{Transmission latency becomes significant when capture latency is improved with better orbit design.
    }
    \label{fig:eval:tx_latency_not_negligible}
\end{figure}





\mybox{
Takeaway: To reduce capture latency, our guidance for operators is to use more planes.
}

\subsection{Constraints of Different Orbital Configurations}
\change{
Altering orbital configurations can impact launch expenses due to varying fuel requirements. 
According to the rideshare program provided by SpaceX~\cite{SpaceX-Rideshare} and Ridespace~\cite{ridespace-Rideshare}, total launching costs are proportional to the satellite count, with payload mass being the dominant factor.
There is also a cost associated with every launch for time/hardware coordination and paperwork.
Assuming that all satellites have similar weights, we model the total launch cost as follows:
$
\text{Cost} = \text{C1} \times \text{Number of Satellites} + \text{C2} \times \text{Number of Launches}
$, 
where $\text{C1} \gg \text{C2}$.

Using more planes increases launch costs by a small portion, since each launch can deploy to only one plane. 
However, companies like Planet and Starlink typically use multiple launches~\cite{planet2017commissioning-smallsat} for their satellite deployments. For example, Planet's recent launch of 36 satellites~\cite{planet-launch-36sat} brings their constellation to 160 satellites.
This incremental deployment strategy accelerates design and manufacturing iterations, enabling rapid improvement with each launch. 
}


Another potential constraint is the applicability of using specific orbits given the federal regulations. A detailed examination of this topic is beyond the scope of this paper. However, given that Starlink satellites operate across various orbital planes and inclinations, it appears feasible to employ multiple planes for Earth observation satellites. Furthermore, SpaceX regularly launches Starlink satellites and offers a rideshare program~\cite{SpaceX-Rideshare}, enabling Earth observation nanosatellites to leverage these launches to achieve desired orbital configurations.

\section{Reducing Transmission Latency}
\label{sec:tx_latency}

After choosing a better orbit design to reduce the capture latency, the transmission latency is no longer negligible, as shown in
\autoref{fig:eval:tx_latency_not_negligible} : it accounts for $56 \%$ of the $95^{th}$ percentile end-to-end latency.
In this section, we consider how to use geo-distributed ground stations to reduce transmission latency.

\begin{figure}[t]
    \centering
    \includegraphics[width=0.8\linewidth]{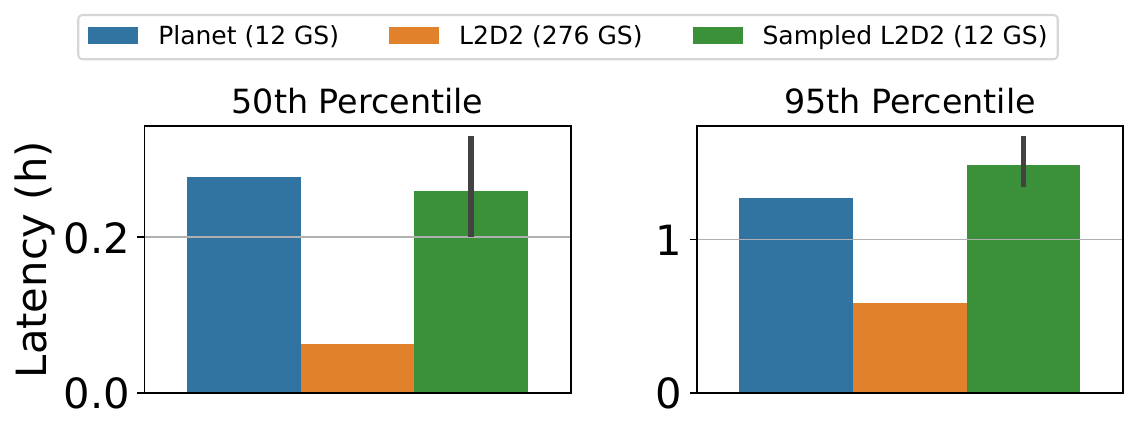}
    \caption{
    Randomly choosing geo-distributed ground stations does not reduce transmission latency.
    }
    \label{fig:eval:tx_latency_sample_motivation}
\end{figure}

\subsection{Strawman Solution}

L2D2~\cite{l2d2-sigcomm} uses 276 geo-distributed SatNOGS ground stations to reduce transmission latency. However, deploying and managing such a large number of ground stations would significantly increase the cost. As a strawman solution, we consider using only a randomly selected subset of SatNOGS to reduce costs. We randomly select 12 SatNOGS ground stations for 30 times. As shown in \autoref{fig:eval:tx_latency_sample_motivation}, we find that randomly selecting ground stations does not reduce and sometimes increases transmission latency compared to the Planet ground stations. This is due to the ground station coverage overlapping as explained below.

\begin{figure}[t]
    \centering
    \includegraphics[width=0.5\linewidth]{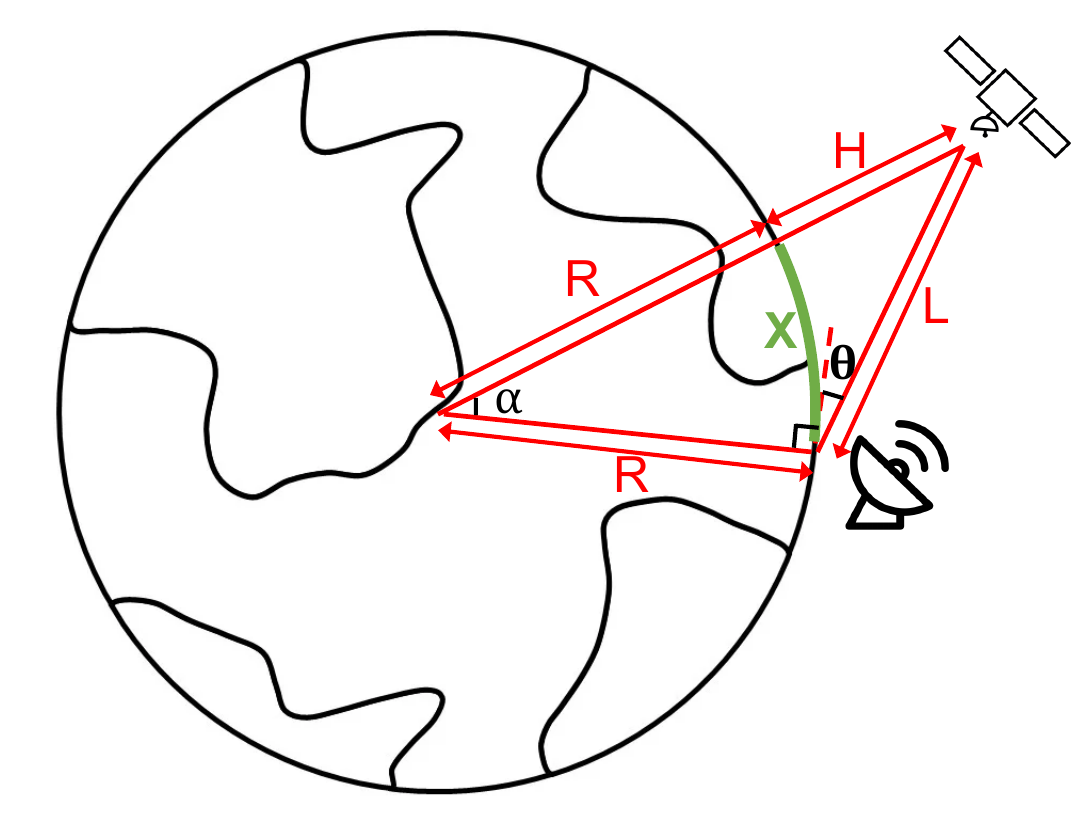}
    \caption{A satellite can communicate with a ground station when the elevation angle ($\theta$) exceeds a minimum threshold.}
\label{fig:tx:gs_coverage_explain}
\end{figure}

\begin{figure}[t]
    \centering
    \includegraphics[width=0.8\linewidth]{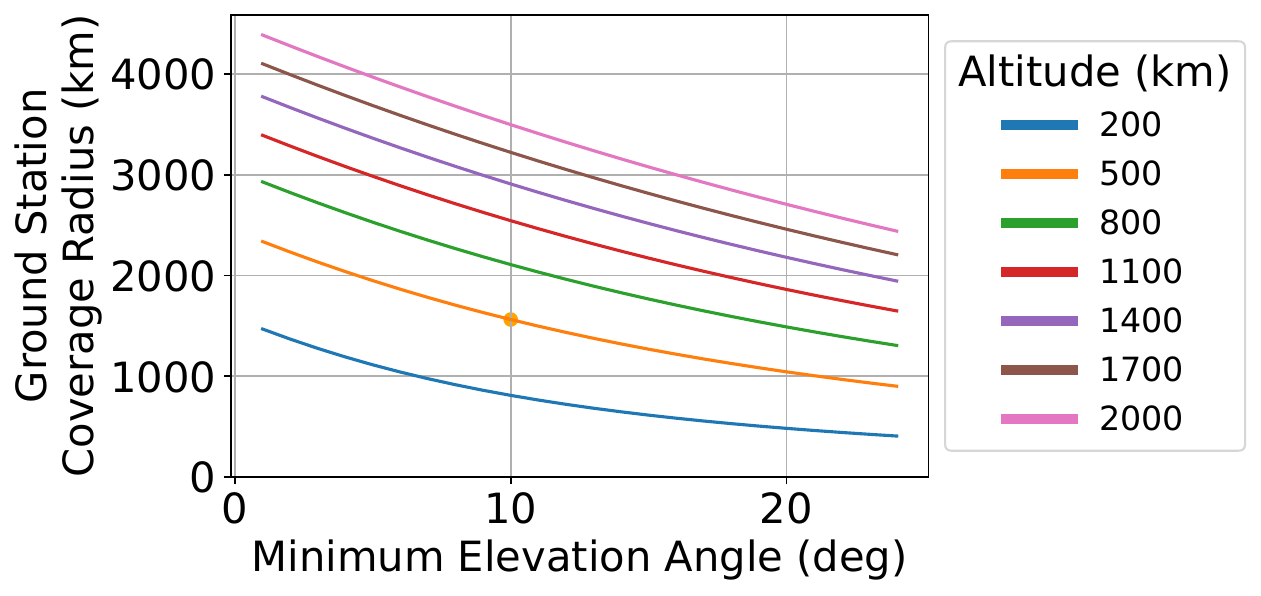}
    \caption{Ground stations typically have wide communication coverage}
\label{fig:tx:gs_coverage_expression}
\end{figure}

\begin{figure}[t]
    \begin{subfigure}[b]{0.48\linewidth}
        \centering
        \includegraphics[width=0.9\linewidth]{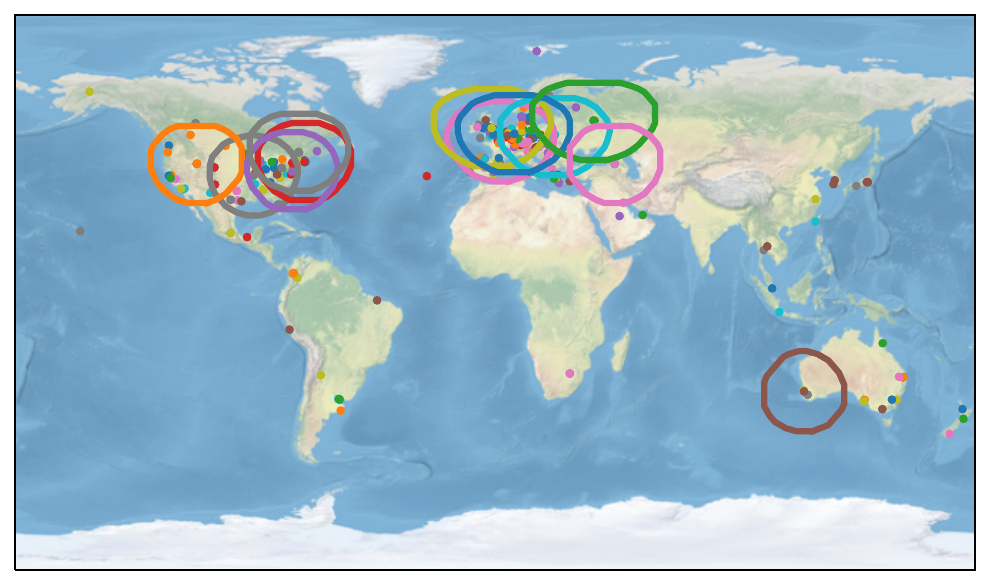}
        \caption{}
    \end{subfigure}
    \begin{subfigure}[b]{0.48\linewidth}
        \centering
        \includegraphics[width=0.9\linewidth]{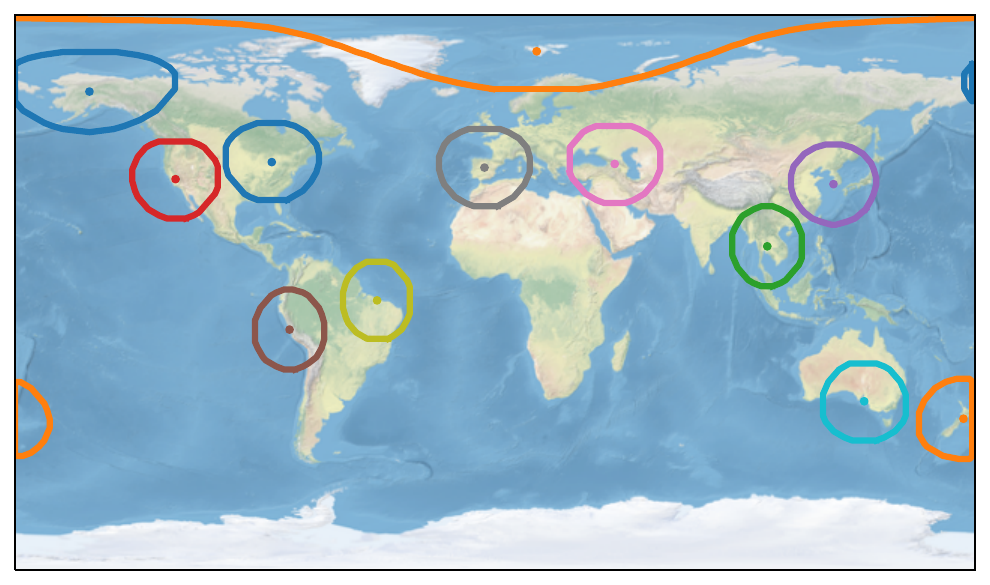}
        \caption{}
    \end{subfigure}
    \caption{(a) Each ground station has a large coverage. Randomly selecting 12 ground stations results in many coverage overlapping. (b) \SYS choose 12 ground stations with minimal overlapping to maximize the coverage.}
    \label{fig:tx:vis_gs_large_coverage}
\end{figure}

\begin{figure}[t]
    \centering
    \includegraphics[width=0.4\linewidth]{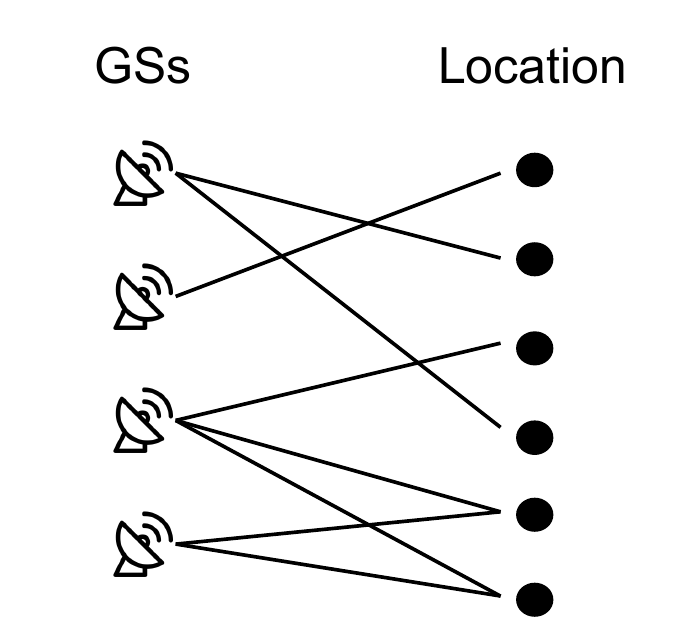}
    \caption{\SYS selects a set of
ground stations that cover most locations.}
\label{fig:tx:tx_design_gs_cover}
\end{figure}

As shown in \autoref{fig:tx:gs_coverage_explain}, a satellite can communicate with a ground station when the elevation angle ($\theta$) between them exceeds a minimum threshold $\theta_{\text{min}}$ (e.g., 10 degrees). 
The ground station coverage
\footnote{Coverage means a satellite can communicate with the ground station when the satellite's projection point on the ground is within the coverage area.}
is a circle with radius $X$, depending on the satellite altitude $H$ and the elevation angle $\theta_{\text{min}}$.
We derive $X$ by solving the triangle: using the law of cosines, \((R + H)^2 = L^2 + R^2 - 2LR \cos(90^\circ + \theta_{\text{min}})\); using the law of sines, \(\frac{L}{\sin(\alpha)} = \frac{R + H}{\sin(90^\circ + \theta_{\text{min}})}\); and for arc length, \(X = R \alpha\). We solve these equations and plot $X$ with different $H$ and $\theta_{\text{min}}$ in \autoref{fig:tx:gs_coverage_expression}.

We find that ground stations typically have wide communication coverage. For example, with $\theta_{\text{min}} = 10^\circ$ and a satellite at 500 km altitude, the ground station coverage radius is 1500 km.
\autoref{fig:tx:vis_gs_large_coverage} illustrates the ground station locations (dots) and their coverage areas (circles).
\autoref{fig:tx:vis_gs_large_coverage}(a) shows that randomly selecting 12 of the 276 SatNOGS ground stations would result in
overlapping coverage areas, which would waste resources.
As comparison, \autoref{fig:tx:vis_gs_large_coverage}(b) shows the 12 ground stations selected by \SYS.

\subsection{Our Approach}
\label{ssec:tx_select_alg}

To maximize overall coverage, we aim to select ground stations with minimal overlap. Since ground stations must be deployed in specific regions to ensure reliable backhaul internet, we propose an algorithm that selects locations based on potential sites (e.g., SatNOGS) and maximizes coverage. Our approach is also applicable to the ground station rental model (e.g., AWS Ground Station~\cite{amazon-ground-station}, Azure Orbital Ground Station~\cite{Azure-Ground-Station}, KSAT~\cite{ksat-ground-station}), where satellite operators choose which ground stations to rent.

%
The input contains the locations of the potential ground stations $GS = \{ gs_i \mid i = 1 \text{ to } n \}$, the budget to select $k$ ground stations, and the event locations $E = \{ e_j \mid j = 1 \text{ to } m \}$.
We consider events to be uniformly distributed throughout the world and discuss how to optimize toward events in a specific place in \autoref{ssec:optimize_specific_loc}. We divide the globe into a uniform grid~\cite{revisit-interval-sentinel-2017} and consider each cell central point as an event location. 
The output is the k selected ground stations.

The algorithm has two steps.
The first step is to calculate a coverage matrix A(n * m) where A[i,j] = 1 indicates that $gs_i$ covers
$e_j$, and 0 means that it does not.
Based on the coverage matrix, we construct a cover graph as shown
in \autoref{fig:tx:tx_design_gs_cover}.  The second step is to find the best
set of ground stations that cover the maximum number of events.  
We formulate
it as a set cover problem and use Integer Linear Programming (ILP) to find an
optimal solution.
The ILP variables are the ground station selection index $s_i$, which indicates if $gs_i$ is selected (1) or not (0), and the event covered index $c_j$, which indicates if event $j$ is covered (1) or not (0).
There are two constraints. The first constraint ensures that $k$ ground stations are selected in total:
$$
\sum_i s_i = k$$
The second constraint ensures that an event is covered by at least one ground station if the covered index is 1:
$$
\forall j, \sum_i (A[i,j] \cdot s_i) \geq c_j$$
The goal is to maximize the number of events covered:
$$\textit{Goal:\ } \ \max \sum_j c_j$$


\begin{figure}[t]
    \centering
    \includegraphics[width=0.97\linewidth]{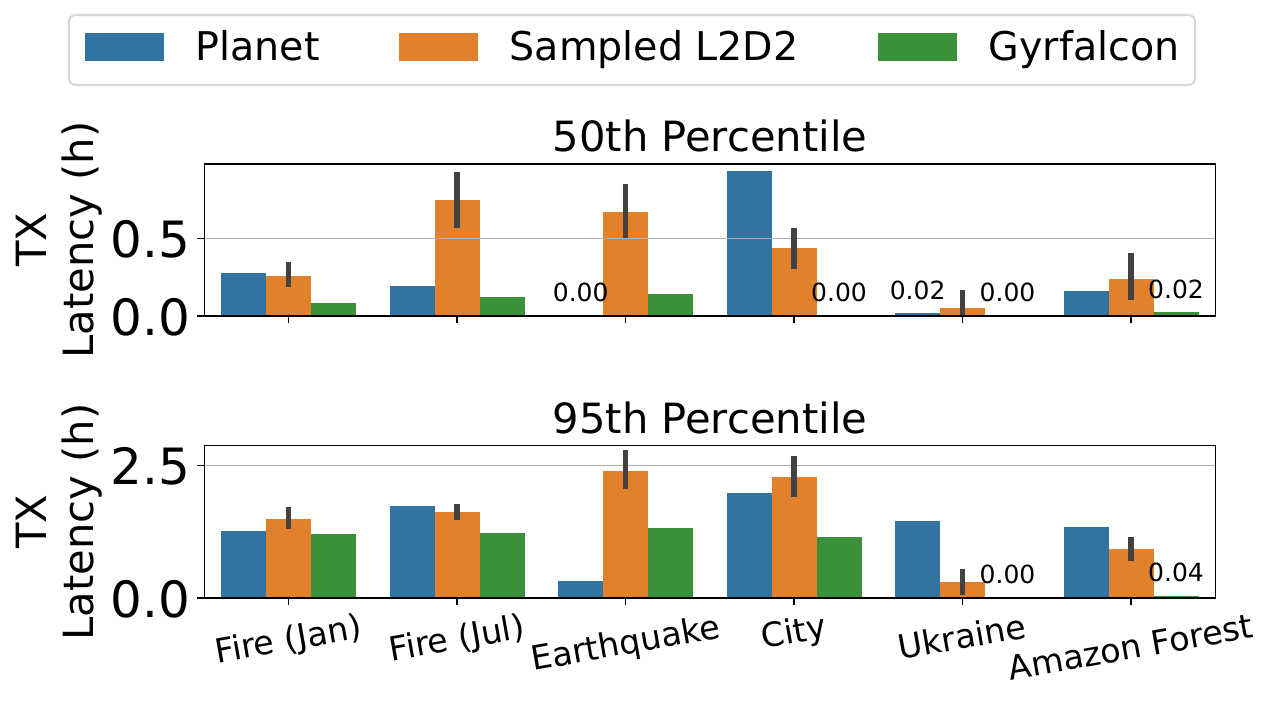}
    \caption{
    \textbf{Transmission latency across six use cases in different systems.}
    \SYS reduces transmission latency compared to Planet and Sampled L2D2.
    }
    \label{fig:tx:all}
\end{figure}

\autoref{fig:tx:all} shows that \SYS reduces transmission latency by $4.9\times$ and $7.7\times$
compared to Planet and randomly-sampled L2D2.  Randomly sampled L2D2 often has
higher latency than Planet.
Certain cases see very low transmission latency (e.g., Earthquake,
Amazon Forest, Ukraine) because these cases have ground stations clustered near
events, covering most events immediately.

\begin{figure}[t]
    \centering
    \includegraphics[width=0.97\linewidth]{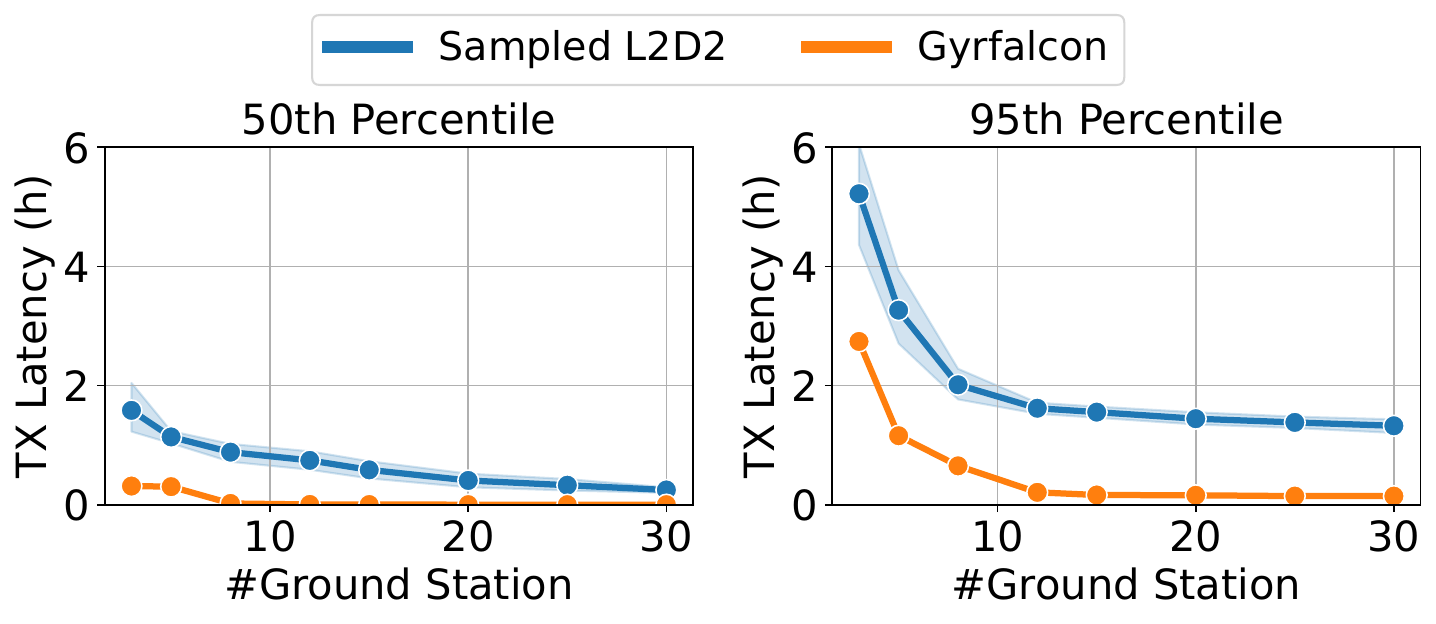}
    \caption{
    \textbf{Transmission latency with different numbers of ground stations.}
    More ground stations reduces transmission latency. 
    \SYS further reduces latency compared to sampled L2D2.}
    \label{fig:tx:smart_select_gs_num}
\end{figure}

\begin{figure}[t]
    \centering
    \includegraphics[width=0.97\linewidth]{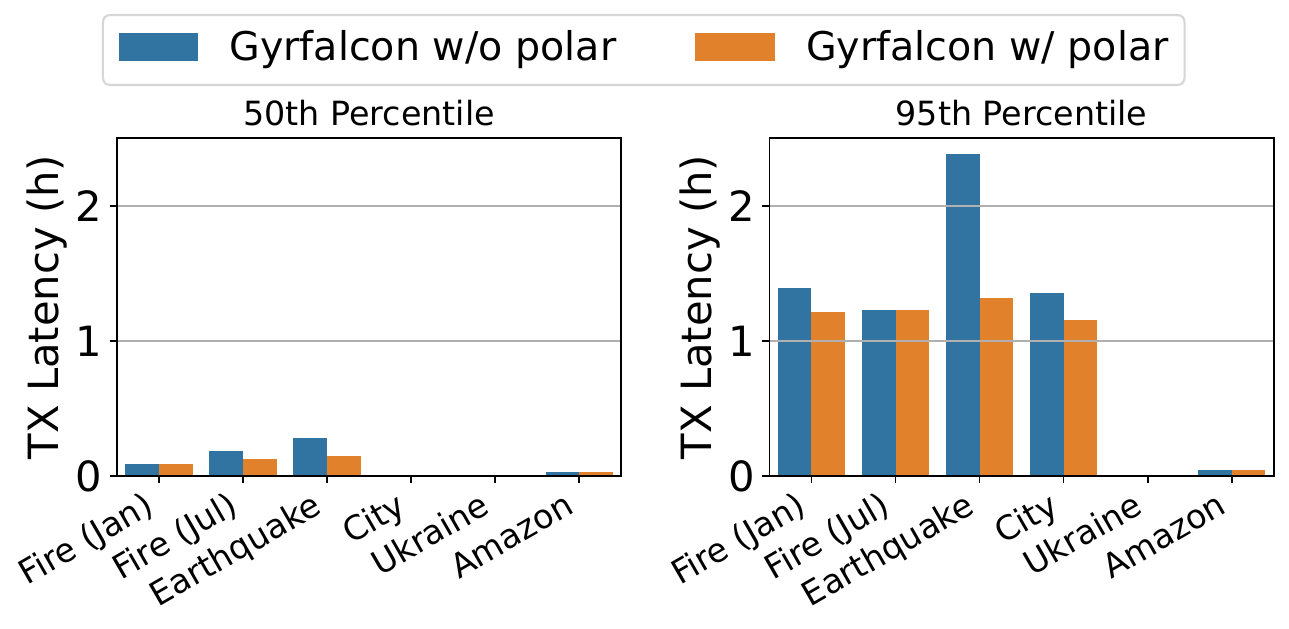}
    \caption{
    \textbf{Transmission latency with and without manually selecting a polar ground station.}
    Manually selecting a polar ground station reduces the transmission latency.}
    \label{fig:tx:polar}
\end{figure}

%
\autoref{fig:tx:smart_select_gs_num} shows that
as the number of ground stations increases from one to 30, transmission latency decreases. 
Median transmission latency decreases from nearly two hours to a few minutes with 30 ground stations.  Tail transmission latency varies from about five hours to less than half an hour.  Consistently across the range of ground station counts, \sys has significantly lower transmission latency than L2D2.
The data also exhibit a trend of diminishing returns as the number of ground stations increases.

\subsection{Advantage of polar ground stations}

Commercial deployments, such as Planet~\cite{Planet-homepage}
and KSat~\cite{ksat-ground-station} strategically place ground stations near the
poles (\eg Svalbard, Antarctica) to ensure that satellites in polar orbits have at least one contact per orbital period. In contrast, ground stations located at mid-latitudes or near the equator may require satellites to complete multiple orbits before establishing contact.

Our algorithm (\autoref{ssec:tx_select_alg}) selects ground stations far apart, but does not guarantee the inclusion of a polar ground station. We evaluate whether manually adding a polar ground station (setting $s_i=1$ for a polar ground station) reduces transmission latency. 
\autoref{fig:tx:polar} demonstrates that incorporating a polar ground station reduces median latency by up to $1.94\times$ and the $95^{th}$ percentile latency by up to $1.8\times$. This data indicates that designer expertise can enhance the optimization results generated by \sys.

\subsection{How orbital configuration affect transmission latency}

\begin{figure}[t]
    \centering
    \includegraphics[width=0.9\linewidth]{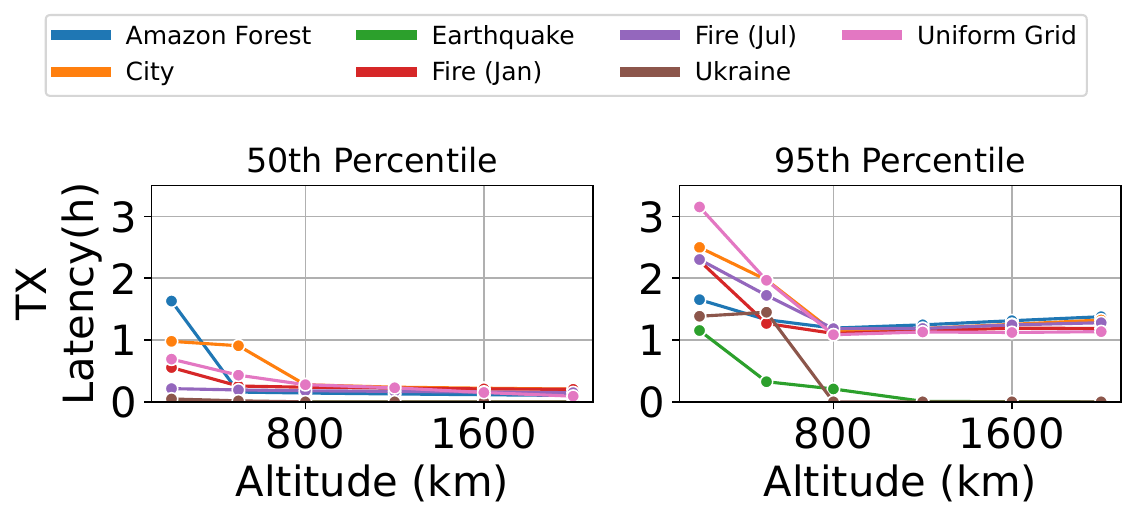}
    \caption{
    Higher altitudes reduce transmission latency.
    }
    \label{fig:eval:tx_latency_altitude}
\end{figure}

\begin{figure}[t]
    \centering
    \includegraphics[width=0.9\linewidth]{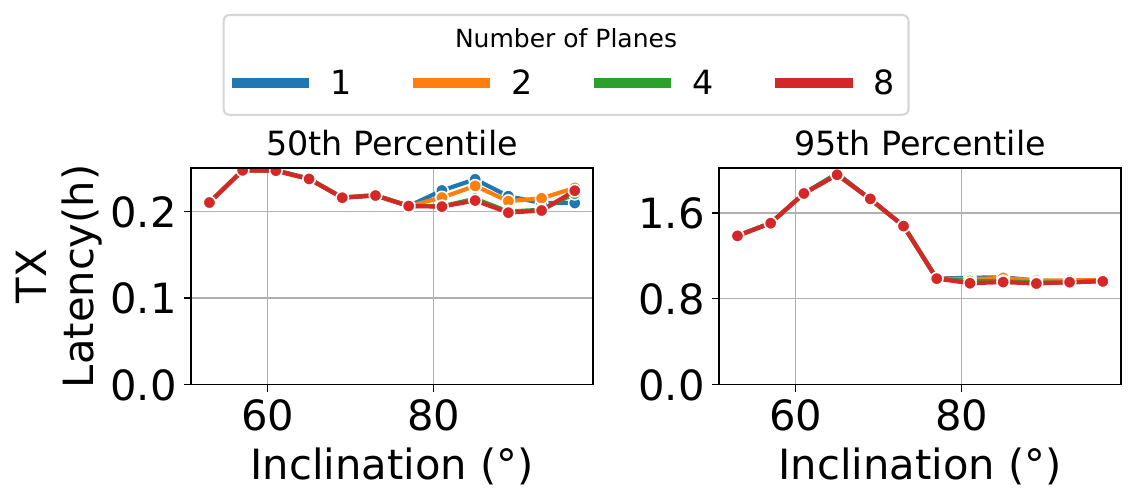}
    \caption{
    Transmission latency when using different orbital configurations.
    }
    \label{fig:eval:tx_orbit}
\end{figure}

\change{
\mypara{Altitude} 
\autoref{fig:eval:tx_latency_altitude} shows that higher altitudes reduce transmission latency. This is because a satellite can communicate with a ground station when its elevation angle exceeds a certain threshold. With a fixed threshold angle, the ground station's coverage area increases as the satellite's altitude rises.
However, as mentioned in \autoref{ssec:orbital_para}, altitude also affects resolution and mission life and is constrained by federal regulations. Therefore, we consider selecting the optimal altitude to be outside the scope of this paper.

\mypara{Other Parameters}
We find that other orbital parameters have a smaller effect on transmission latency. For brevity, we present results only for inclination and the number of planes.
As shown in \autoref{fig:eval:tx_orbit}, using more planes barely reduces transmission latency and using lower inclination even increases the latency.
Considering that capture latency is significantly greater than transmission latency and that the number of planes and inclination have a larger impact on capture latency, we conclude that operators should prioritize optimizing capture latency when selecting orbit configurations.
}

\section{Discussion}
\label{sec:discuss}

We show that \sys can be used for both fresh and incremental deployments. We also show how to optimize the constellation and ground stations to target specific event locations.

\subsection{Incremental Deployment}


\begin{figure}[t]
\centering
\begin{subfigure}[b]{0.9\linewidth}
    \centering
    \includegraphics[width=0.97\linewidth]{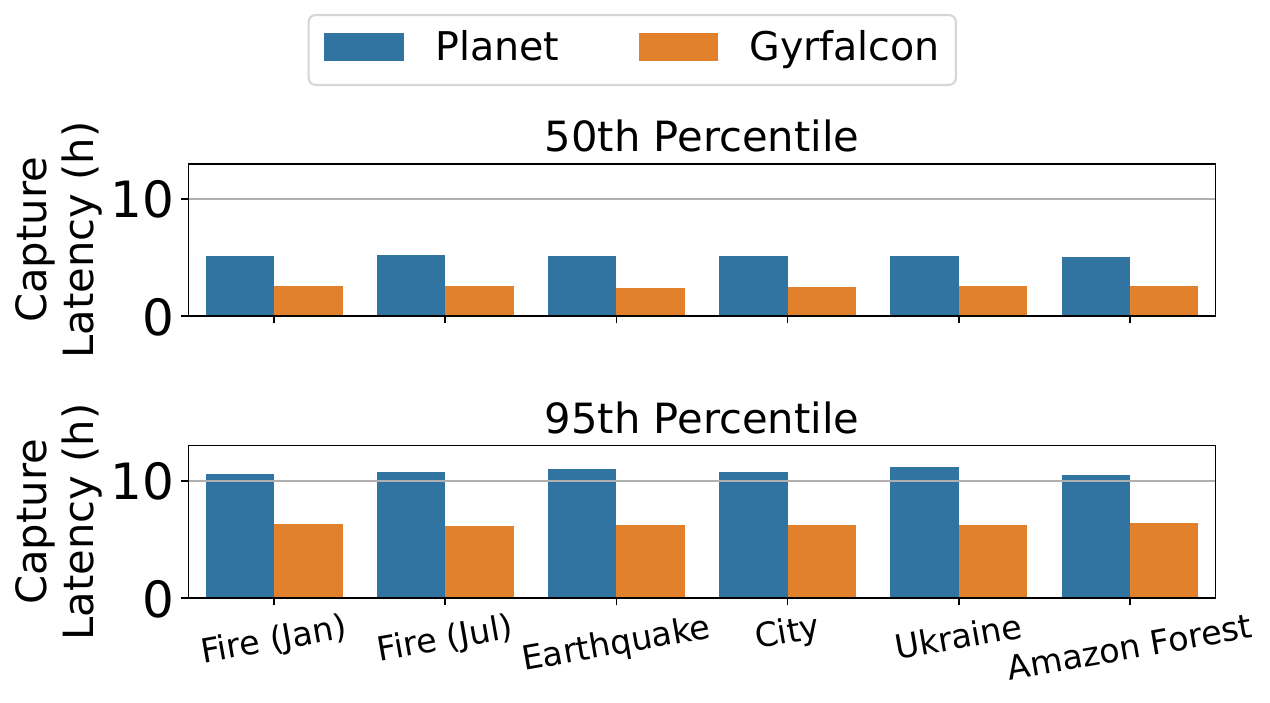}
    \caption{Capture Latency}
\end{subfigure}
\begin{subfigure}[b]{0.9\linewidth}
    \centering
    \includegraphics[width=0.97\linewidth]{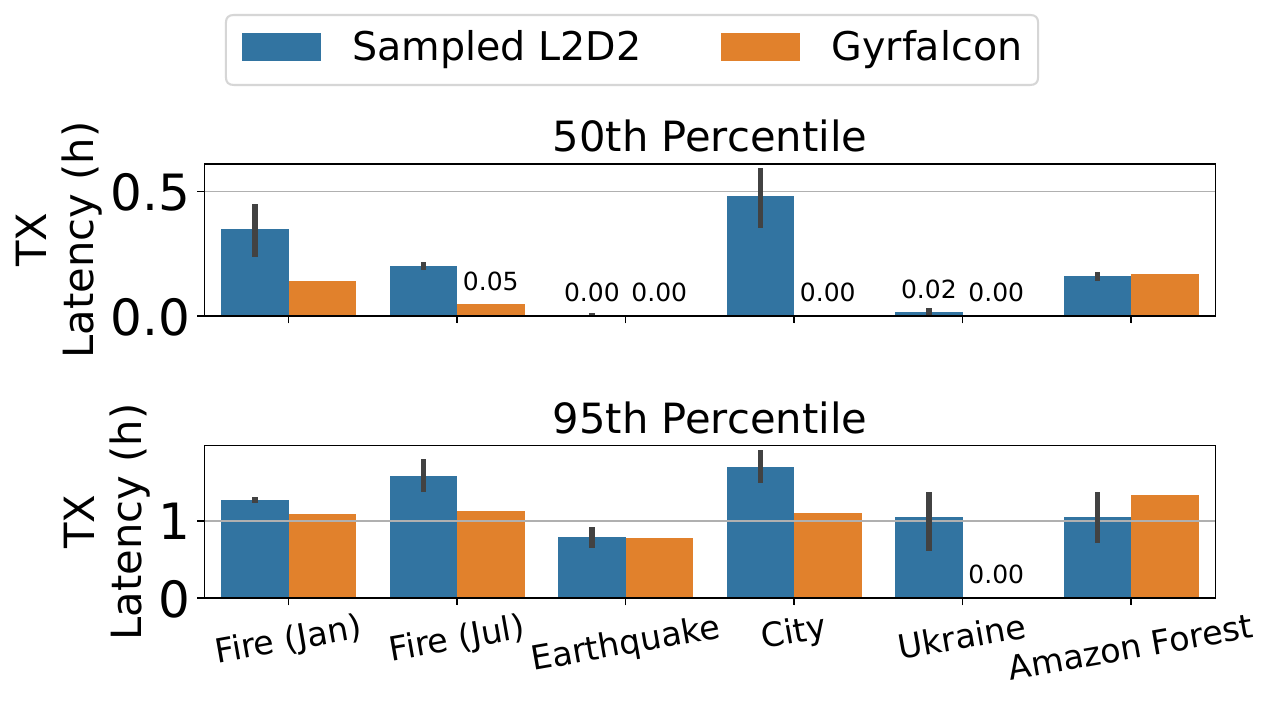}
    \caption{Transmission Latency}
\end{subfigure}
\caption{
For incremental deployment, \SYS reduces capture and transmission latency compared to the existing system. 
}
\label{fig:eval:incremental}
\end{figure}

We evaluated \SYS's ability to support incremental constellation deployment.
Planet's latest launch includes 36 satellites~\cite{planet-launch-36sat}, so we examine whether \SYS can
reduce capture latency if those 36 satellites were instead launched
incrementally into other orbits.  We also consider the deployment of three
additional ground stations and compare the transmission latency of
\SampledLLDD\ and \SYS.
\autoref{fig:eval:incremental} shows that \SYS reduces the capture latency by
\evalIncCapture and the transmission latency by \evalIncTX.
\autoref{fig:eval:orbit_visual}(c)(d) shows the incremental deployment constellation.

\subsection{Optimizing for Specific Locations}
\label{ssec:optimize_specific_loc}


\begin{figure}[t]
    \centering
    \includegraphics[width=0.97\linewidth]{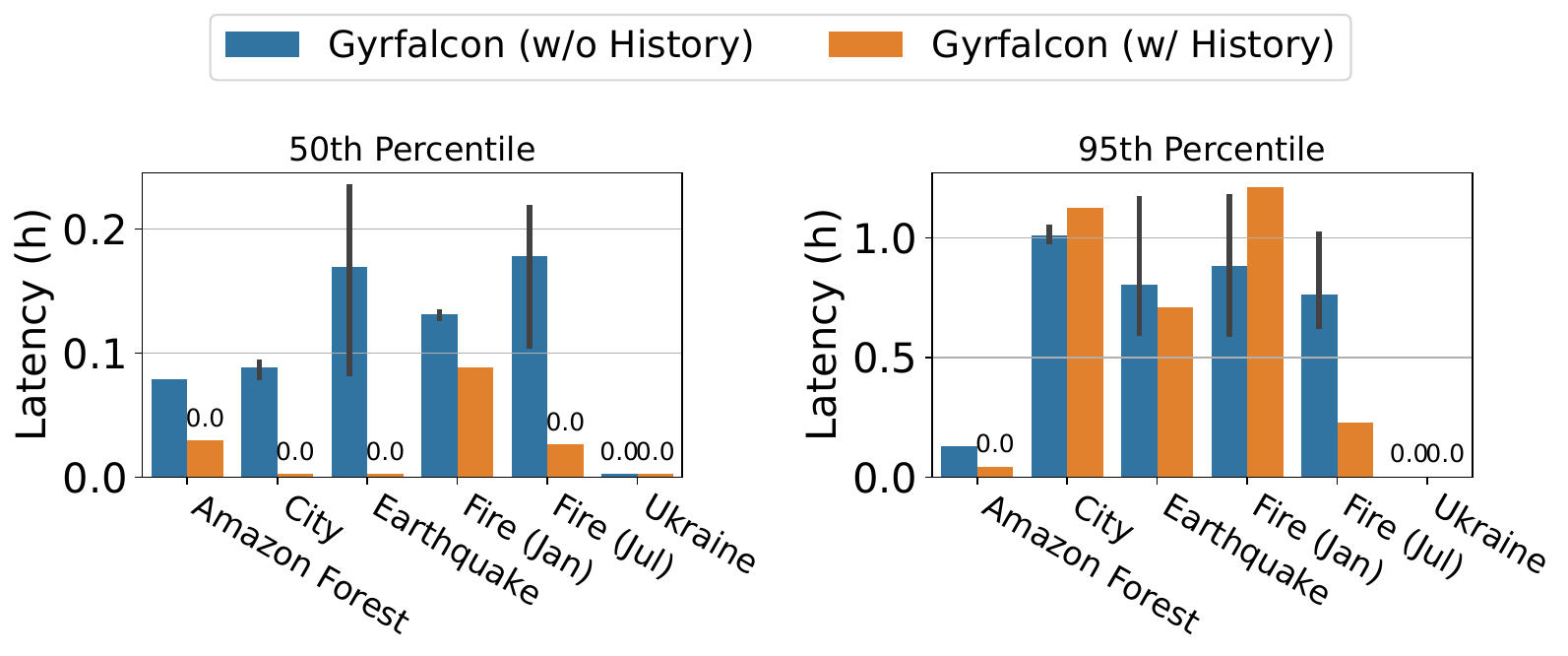}
    \caption{
    \textbf{Choosing ground stations near historical event sites reduces transmission latency.}}
    \label{fig:tx:history_benchmark}
\end{figure}

\change{
For capture latency, operators can use a lower inclination orbit to reduce the latency for events at a lower latitude. But, as discussed earlier, it might miss events at a higher latitude.

For transmission latency, if operators know the potential event location beforehand (e.g., monitoring forest fires in California, tracking wildlife in Ukraine), they can deploy a ground station nearby the event to achieve zero transmission latency. The input to the algorithm in \autoref{ssec:tx_select_alg} will then change $E$ from uniformly distributed events to specific locations derived from historical events or the application's regions of interest. 
\autoref{fig:tx:history_benchmark} demonstrates that selecting ground stations near historical event sites can further reduce transmission latency. We combined various event locations to determine the optimal ground station placement, ensuring the system's adaptability across different scenarios. However, for certain events, tail latency may increase due to imperfect predictions based on historical data.

Note that since each ground station has a large communication coverage (e.g., a 3000 km diameter circle), it does not need to be placed right next to the event location. This flexibility allows operators to choose ground station sites with lower costs and better backhaul connections.
}

\subsection{Compute Latency}
\label{ssec:eval:compute}
\begin{figure}[t]
   \centering
   \includegraphics[width=0.9\linewidth]{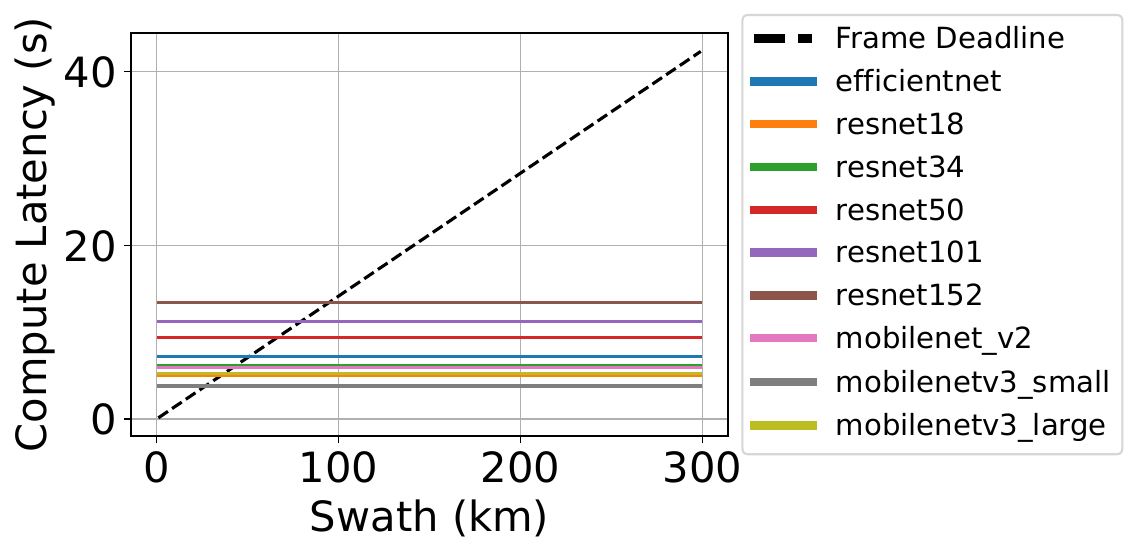}
   \caption{Compute Latency and Frame Deadline.
   }
   \label{fig:eval:compute_latency}
\end{figure}

We consider compute latency as delay between an image's capture and availability
for downlink if an event is detected in the image.
Compute latency for different ML models is not an impediment to \SYS with
reasonable swath width.  We studied a range of swaths and models (\autoref{fig:eval:compute_latency}) and found
that, while frame deadline increases with larger swath, with swath greater than
\SI{100}{km}, all models complete processing within the frame deadline. 
A wide swath has high coverage and low capture latency and we reasonably do not
consider scenarios where compute latency exceeds frame deadline.
Prior techniques on model compression~\cite{hinton2015distilling},
quantization~\cite{han2016deep}, and tiling~\cite{kodan-asplos23} reduce
compute latency and are complementary.

\subsection{New Sensor and Communication Technology}
New sensors, such as wider swath cameras, hyperspectral cameras, or SAR provide a larger swath. However, as mentioned in ``Discussion on Swath'' in \autoref{ssec:orbital_para}, our design on orbital configuration provides an orthogonal way to reduce capture latency. 
\autoref{fig:capture:planet_large_swath} shows that a better design reduce capture latency across different camera swath from 5km to 125km.

Another promising direction is to use crosslink~\cite{hotnet20-without-ISL, network-topology-design-conext19-eth} between satellites to transmit data to reduce latency.
However, these works consider large satellites (\eg Starlink satellites weighting around $227-1250$ kg) with more power.
To the best of our knowledge, no existing nanosatellite deployment supports long-distance crosslink communication. Our work provides a practical way to choose the optimal location of the ground station to reduce communication latency.

\change{
\subsection{Sun-synchronous Orbits}

Earth-observing satellites are typically placed in sun synchronous orbits, ensuring they pass over the Earth's surface at the same local mean solar time, maintaining consistent illumination, which will simplify image processing. Each altitude has a specific inclination that satisfies the Sun-synchronous orbit requirement.
Previously, we demonstrated that the most crucial factor in improving capture latency is using multiple planes. This approach is compatible with Sun-synchronous orbits, as all these planes can still maintain Sun-synchronous characteristics.

\subsection{Cloud Cover on Satellite Images}

Satellite images can be occluded by clouds, making them unusable for downstream event detection. We do not incorporate cloud effects into our simulation for two reasons. First, if a satellite image is cloud-covered, it will be unusable by both baseline systems and \SYS. However, \SYS's constellation and ground station design suggestions remain advantageous for non-cloud images compared to baselines. Second, there are sensors that can penetrate clouds, such as Synthetic Aperture Radar (SAR)~\cite{iceye-sar-satellite,radar-cubesat, 12U-cubesat-sar, 12U-cubesat-sar-snow-ku-band}.
Satellites equipped with these sensors are not affected by cloud cover.
}

\section{Related Work}
Prior work has studied LEO satellite networks,
such as network testbeds~\cite{StarryNet-nsdi23}, network functions
optimization~\cite{SpaceCore-sigcomm22}, studying
vulnerabilities~\cite{ICARUS-attack-atc21-adrian}, internet
routing~\cite{satellite-route-variability-PAM23,
Internet-Backbones-sigcomm20-ccr-eth}, pointing and
maneuvering~\cite{yuanjie-mobicom23-starlink-network}, and orbit design for
network addressing~\cite{yuanjie-hotnets21-cyber} and
performance~\cite{leonet23-network-orbit-design}.
These projects focus on improving communication with a constellation and,
instead, we focus on event detection with the observation constellation.
Recent work on ground stations improves the signal-to-noise ratio, such as
combining signals from multiple ground stations~\cite{Vaibhav-mobicom21-Quasar}
or satellites~\cite{Vaibhav-SelfieStick-ipsn22}, and reconfigurable
metasurfaces~\cite{metasurface-satellite-hotnet22,
liliqiu-metasurface-mobicom23}.
However, their focus is on improving downlink throughput rather than reducing latency.
Umbra~\cite{deepak-umbra-mobicom23} addresses the queueing problem at the ground station. However, we consider that the critical event signal will be given top priority in transmission and queuing delay is negligible. 
Therefore, we are solving a different problem from Umbra.

Orbital edge computing recently emerged for satellite computation.
Kodan~\cite{kodan-asplos23} solves the in-orbit computation bottleneck for
image processing by training specialized ML models for different geospatial
contexts, which can be used to reduce the compute latency for our problem.
%
There are also works that consider other perspectives on orbital edge
computing, such as using orbital edge computing to establish low-latency
networks for applications such as content delivery networks and online
games~\cite{in-orbit-computing-hotnets20-eth, orchestration-sat-hotedge20},
potential failures of orbital edge computing~\cite{oec-fail-survey-leonet23,
hotnet23-space-radiation}, federal learning~\cite{fedspace22} to train ML
models on satellites rather than transmitting images back to Earth for training
to solve the downlink bottleneck.

\section{Conclusion}

We consider using nanosatellite constellations for low-latency critical event detection and show that capture latency is the major component, constituting up to 90\% of end-to-end latency. We analyze how different orbital configurations and ground station locations affect latency, providing deployment guidance for operators.



{
\bibliographystyle{abbrv}
\bibliography{cite/cite,cite/eagleeye,cite/low_latency}
}


\end{document}